\documentclass[prl,twocolumn,preprintnumbers,showpacs,floatfix,amsmath,amssymb]{revtex4}
\usepackage{graphicx}
\usepackage{color}
\usepackage{dcolumn}

\newcommand{\be}{\begin{equation}}
\newcommand{\ee}{\end{equation}}
\newcommand{\ba}{\begin{eqnarray}}
\newcommand{\ea}{\end{eqnarray}}

\newcommand\ie{\textit{i.e.}\ }

\begin{document}

\title{Numerical simulation of ejected molten metal-nanoparticles liquefied
       by laser irradiation: Interplay of geometry and dewetting}

\author{S. Afkhami} 
\email{shahriar.afkhami@njit.edu}
\author{L. Kondic} 
\affiliation{Department of Mathematical Sciences and Center for Applied Mathematics and Statistics, 
             New Jersey Institute of Technology, Newark, NJ 07102 USA}



\begin{abstract}
Metallic nanoparticles, liquefied by fast laser irradiation, go through a rapid change of
shape attempting to minimize their surface energy. The resulting nanodrops may be ejected from
the substrate when the mechanisms leading to dewetting are sufficiently strong, as in the
experiments involving gold nanoparticles [Habenicht et al., Science {\bf 309}, 2043 (2005)].  
We use a direct continuum-level approach to accurately model the process of liquid nanodrop formation
and the subsequent ejection from the substrate.   Our computations
show a significant role of inertial effects and an elaborate interplay of initial geometry and wetting 
properties: e.g., we can control  the direction of ejection by prescribing appropriate initial shape and/or
wetting properties. The basic insight regarding ejection itself can be reached by considering 
a simple effective model based on an energy balance.
We validate our computations by comparing directly with the experiments
specified above  involving the length scales measured in hundreds of nanometers, and with molecular
dynamics simulations on much shorter scales measured in tens of atomic diameters, as in 
M. Fuentes-Cabrera et al., Phys.~Rev.~E {\bf 83}, 041603 (2011).
The quantitative agreement, in addition to illustrating
how to controlling particle ejection, shows utility of continuum-based simulation in
describing dynamics on nanoscale quantitatively, even in a complex setting as considered here.  
\end{abstract}

\maketitle

Evolution of fluid drops deposited on solid substrates
has been a focus of large research effort for decades.   
More recently, this effort has been particularly extensive on the nanoscale, 
due to relevance of nanostructures 
in a variety of fields, ranging from DNA sequencing to
plasmonics~\cite{maier_2007,atwater_natmat10}, nanogmagnetism~\cite{bader_rmp06}, and
liquid crystal displays and solar panel designs~\cite{45}.   
For example, the size and distribution of metallic particles strongly affects the coupling
of surface plasmons to incident energy~\cite{maier_2007}. Controlling this coupling has
the potential for large increases in the yield of solar cell devices.

In addition to physical experiments, modeling and simulations 
provide a significant insight into the effects governing the evolution of drops and other liquid structures.
Numerical simulation has an advantage of allowing to switch off some of the involved
physical phenomena and therefore isolate the dominant ones.   
On  nanoscale, however, it is not trivial to decide on appropriate simulation technique.
Molecular dynamics (MD) simulations, while very powerful and presumably as close to real physical
picture as it is our knowledge of underlying interaction laws, are still extremely computationally 
demanding. Therefore, one would like to resort to the continuum-based
simulation. However, it is not immediately clear that this approach is appropriate on the
nanoscale where the basic assumptions of continuum
fluid mechanics are pushed to their limits.      
In addition, the continuum simulation of free surface time-dependent problems based
on full three-dimensional Navier-Stokes (N-S) formulation is still computationally demanding. 
For this reason, this type of 
simulation is rarely attempted in practice and instead the researchers resort to asymptotic (long-wave)
methods. Such an approach has been applied with
success to problems involving wetting and dewetting of drops and films; 
see~\cite{67} for reviews. This direction is, however, 
questionable due to the inherent assumptions, in 
particular when contact angles are large and inertial effects are significant.   
One such class of problems involves liquid metals. These configurations may even lead to topological
changes under liquefaction with metal particles detaching from the substrate~\cite{habenicht_05}.  In 
such cases, long-wave theory clearly cannot be applied. 

In this work, we show that direct numerical solutions of N-S equations lead to the 
results which are in quantitative agreement with (i) MD simulations of 
dewetting of  $1$-$1.5$ nm thick, liquid copper (Cu) disks of radius of $10$-$15$ nm, 
and (ii) physical experiments involving `jumping' gold (Au) nanoparticles
of typical in-plane length scale in the range of $100$ nanometers.
While utility of continuum mechanics on nanoscale problems involving liquid
metals was discussed previously (see, e.g.,~\cite{Burton04}), 
as far as the authors are aware, this is the first attempt to compare explicitly
continuum N-S simulations both to MD and to physical experiments in a dynamic setting.  
Furthermore, we utilize the simulations to illustrate the 
dominant effects and discuss in particular the conditions leading to ejection of 
fluid material from the substrate. We also present an effective model 
based on an energy balance to provide a basic insight into the ejection mechanism. 
We first consider dewetting of liquid Cu disks recently simulated by 
MD~\cite{fuentes_pre11} and  then shift focus to Au particles considered 
experimentally~\cite{habenicht_05}.   

{\it Methods.}  
The simulations are based on the Volume of Fluid (VoF) approach, which, when coupled to a flow solver, 
is also used to compute quantities related to surface tension that enter the flow calculation.  
Until recently, the VoF method  was not deemed appropriate for the study of phenomena 
in which surface tension is the driving force.  Recent improvements to calculating 
curvature and applying the surface tension force appear to resolve this issue~\cite{VOF}.  
Within this approach, one solves the three-dimensional N-S equations that govern the 
motion of the fluid both inside of the liquid domain, and in the surrounding gas phase. 
For Cu disk structures, a free-slip boundary condition is specified at the substrate; this choice 
is motivated by the MD simulations~\cite{fuentes_pre11}. This assumption is also consistent with the fact that
the length scale associated with slip is nanometric for most systems~\cite{bonn_rmp2009}.
When considering Au structures, the Navier slip \cite{Haley91} with slip length of $3$ nm is imposed; this 
particular value is obtained by direct comparison with the experiments~\cite{habenicht_05} and is further
discussed in Supplementary Materials~\cite{sup_mat}. Such a value is also consistent with 
the MD simulations \cite{Qian04}, indicating slip lengths of a few nanometers for mesoscopic systems.   
The physical problem considered is the one of a Newtonian, isothermal, incompressible
fluid and therefore in the present work we do not consider the effects related to phase 
change and/or thermal variation of material properties occurring in experiments~\cite{habenicht_05}.   
Simulations are carried out on an adaptive mesh \cite{Popinet03}, and numerical convergence has
been verified by considering simulations with different grid resolutions \cite{afkhami_jcp09}.   
See~\cite{sup_mat} for more details.  
\begin{figure}[thb]
\centering
\begin{tabular}{cc}
  \includegraphics[width=0.235\textwidth]{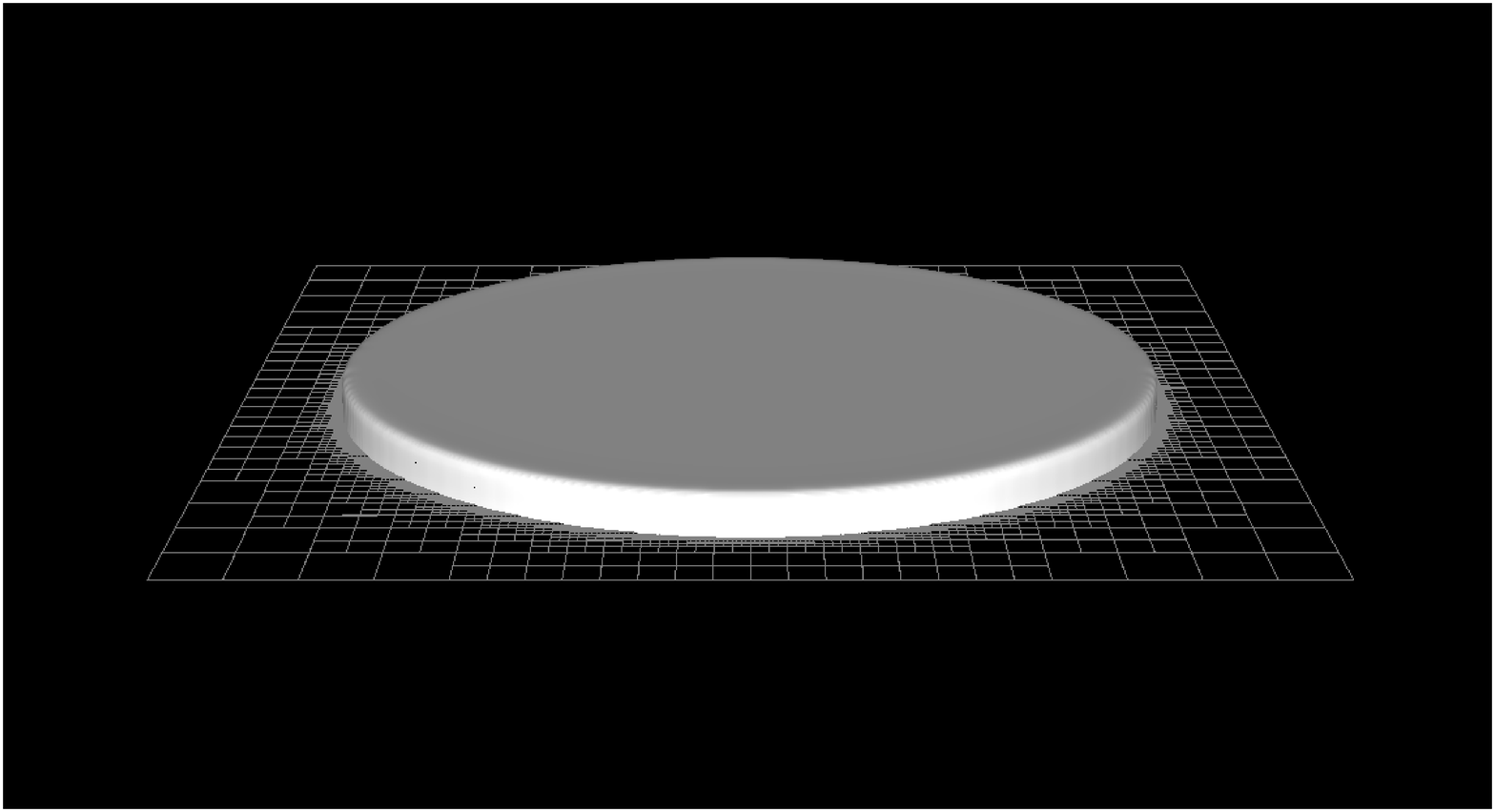}&
  \includegraphics[width=0.235\textwidth]{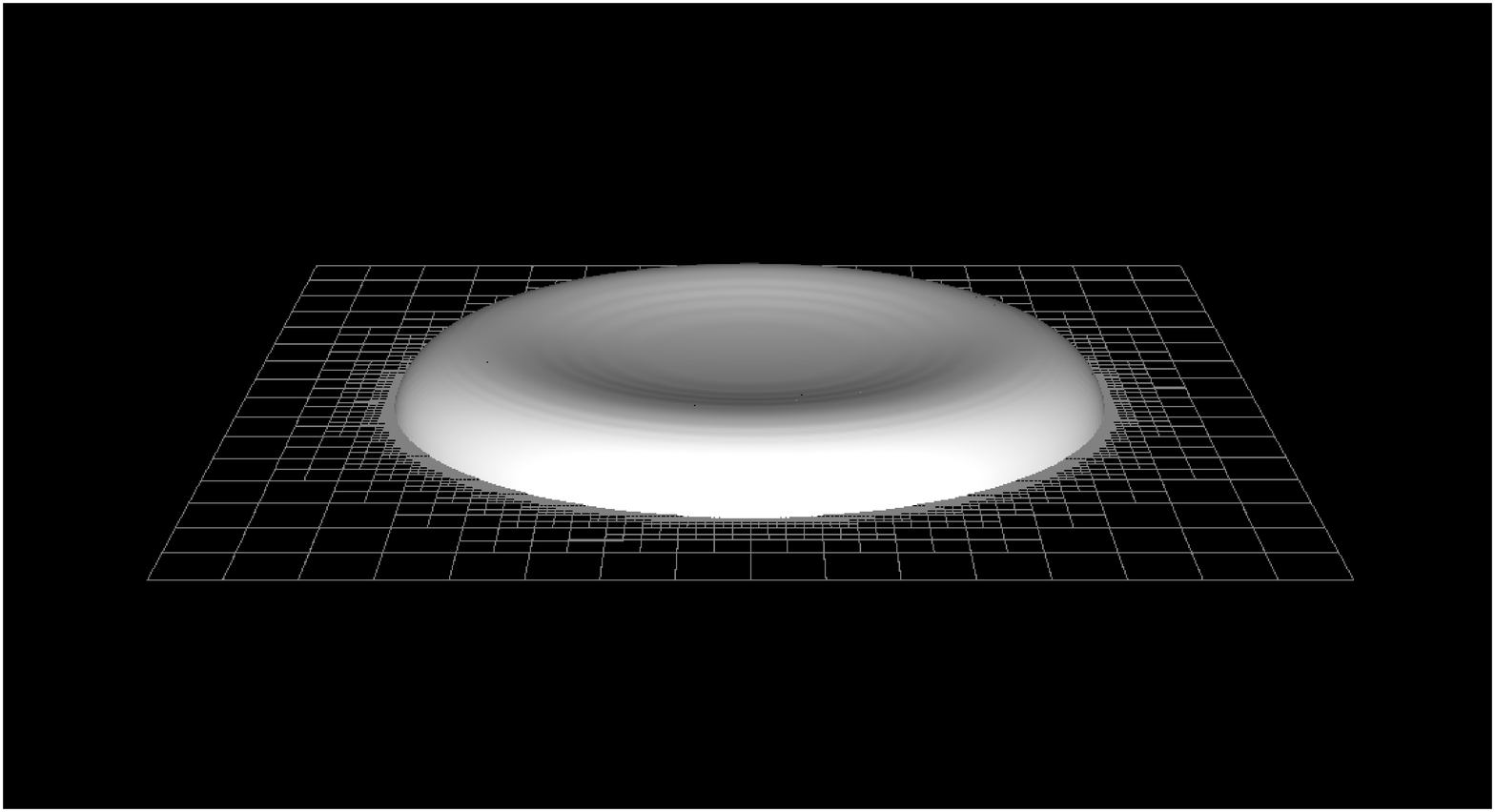}\\  
  0 ps&20 ps\\
  \includegraphics[width=0.235\textwidth]{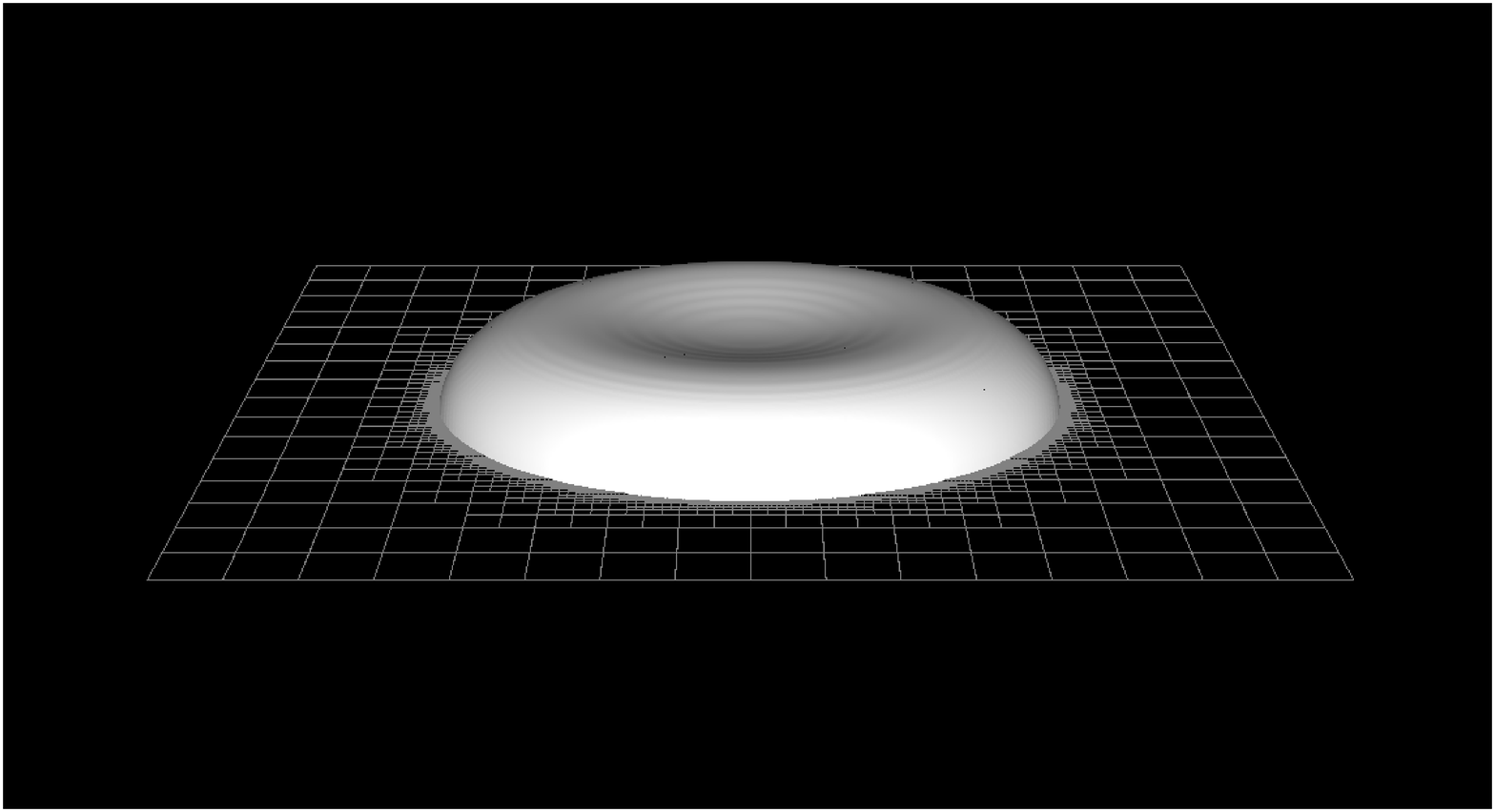}&
  \includegraphics[width=0.235\textwidth]{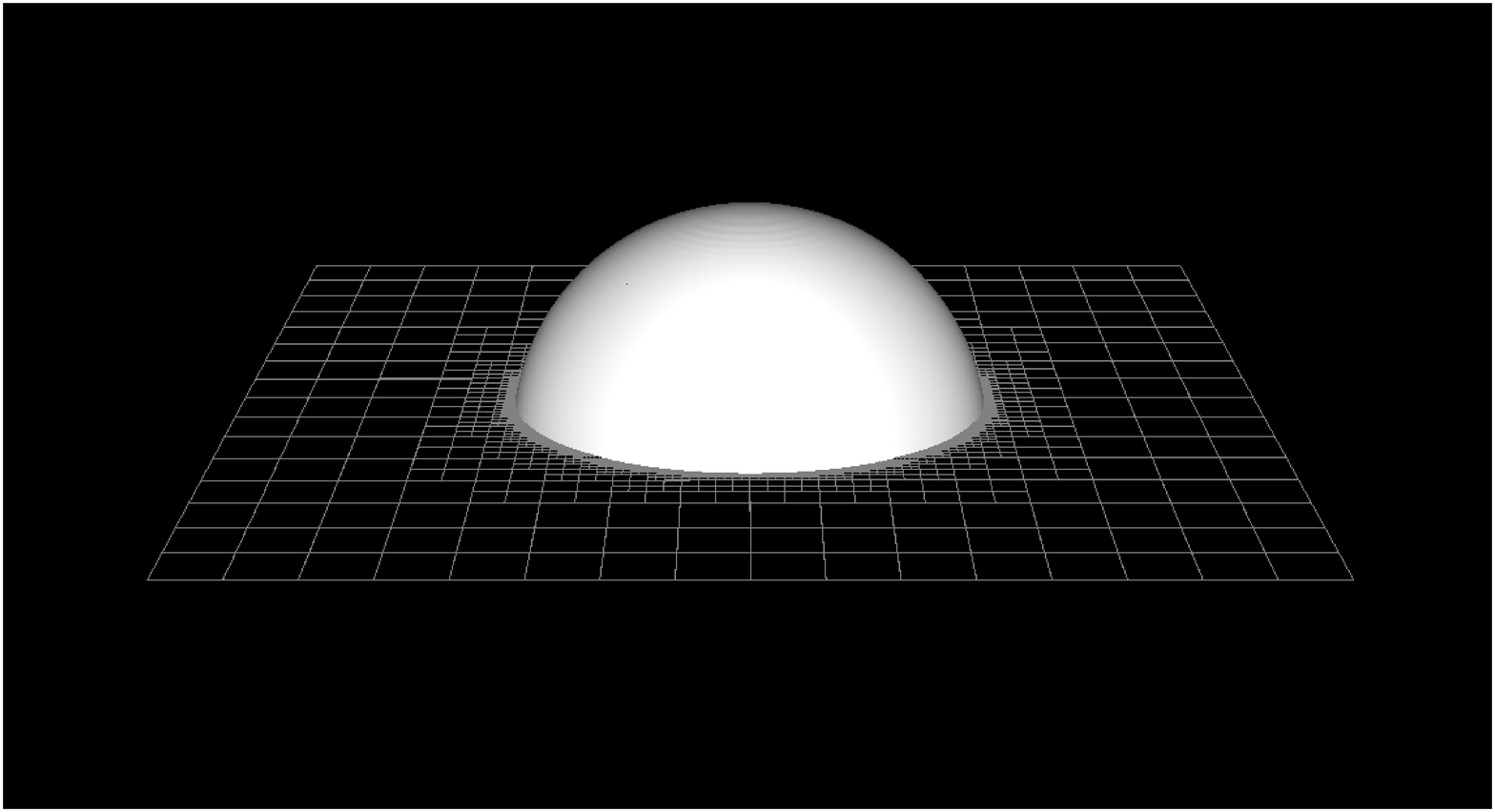}\\  
  30 ps&200 ps\\
&\hspace{-47mm}(a)\\
\includegraphics[width=0.25\textwidth, trim=4mm 2mm 0 0]{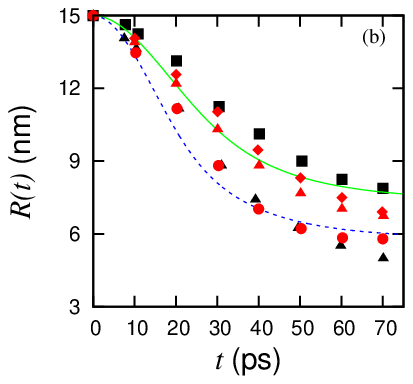} &
\includegraphics[width=0.25\textwidth, trim=4mm 2mm 0 0]{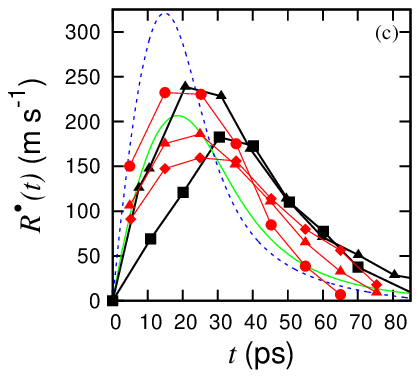} 
\end{tabular}
\caption{(Color online) Evolution of a Cu disk ($h_0=15$ \AA, $R_0=150$ \AA, $\theta_{0}=90^\circ$); 
(a) Snapshots for $\theta_{\text{eq}}=80{^\circ}$;
(b) $R(t)$  and (c) $\dot R(t)$ for ({\small $\blacksquare$}) $\theta_{\text{eq}}=80{^\circ}$
and ($\blacktriangle$) $\theta_{\text{eq}}=115{^\circ}$.
MD simulations for ({\color[rgb]{1.00,0.00,0.00}{$\bullet$}}) LJ-a, $\theta_{\text{eq}}=116{^\circ}$, 
and ({\color[rgb]{1.00,0.00,0.00}{$\blacklozenge$}}) LJ-b,  $\theta_{\text{eq}}=75.6{^\circ}$, with $h_0=10$ \AA~and 
({\color[rgb]{1.00,0.00,0.00}{$\blacktriangle$}}) LJ-b with $h_0=15$ \AA~\cite{fuentes_pre11}.
Predictions of the model, Eq.~(\ref{eq:energy}), for $\theta_{\text{eq}}=80^\circ$ (solid green line)
and $115^\circ$ (dashed blue line). $Oh\approx0.35$.}
\vspace{-0.1in}
\label{fig:3D80}
\end{figure}

\begin{figure}[thb]
\centering
\begin{tabular}{cc}
  \includegraphics[width=0.235\textwidth]{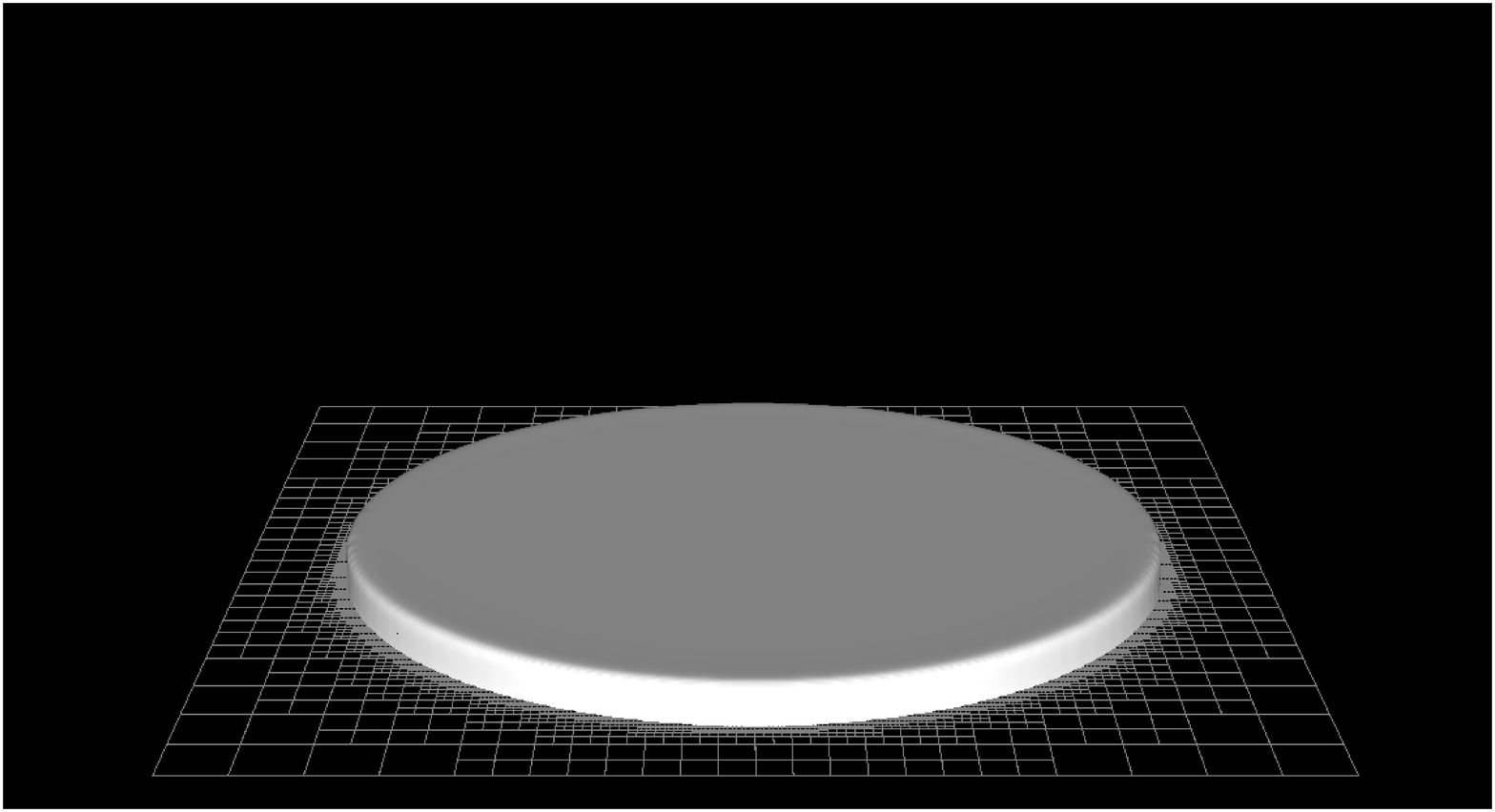}&
  \includegraphics[width=0.235\textwidth]{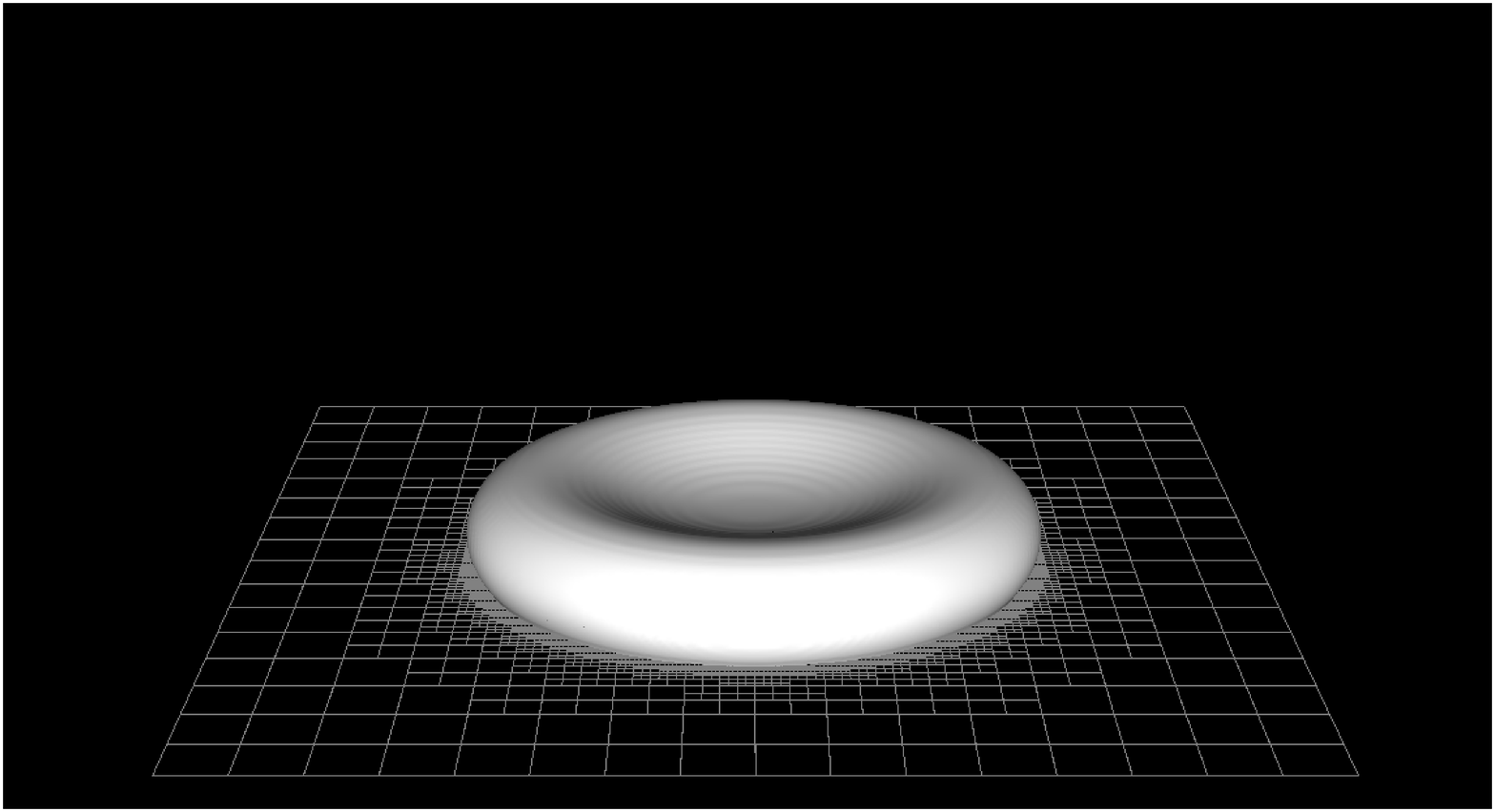}\\  
  0 ps& 20 ps\\
  \includegraphics[width=0.235\textwidth]{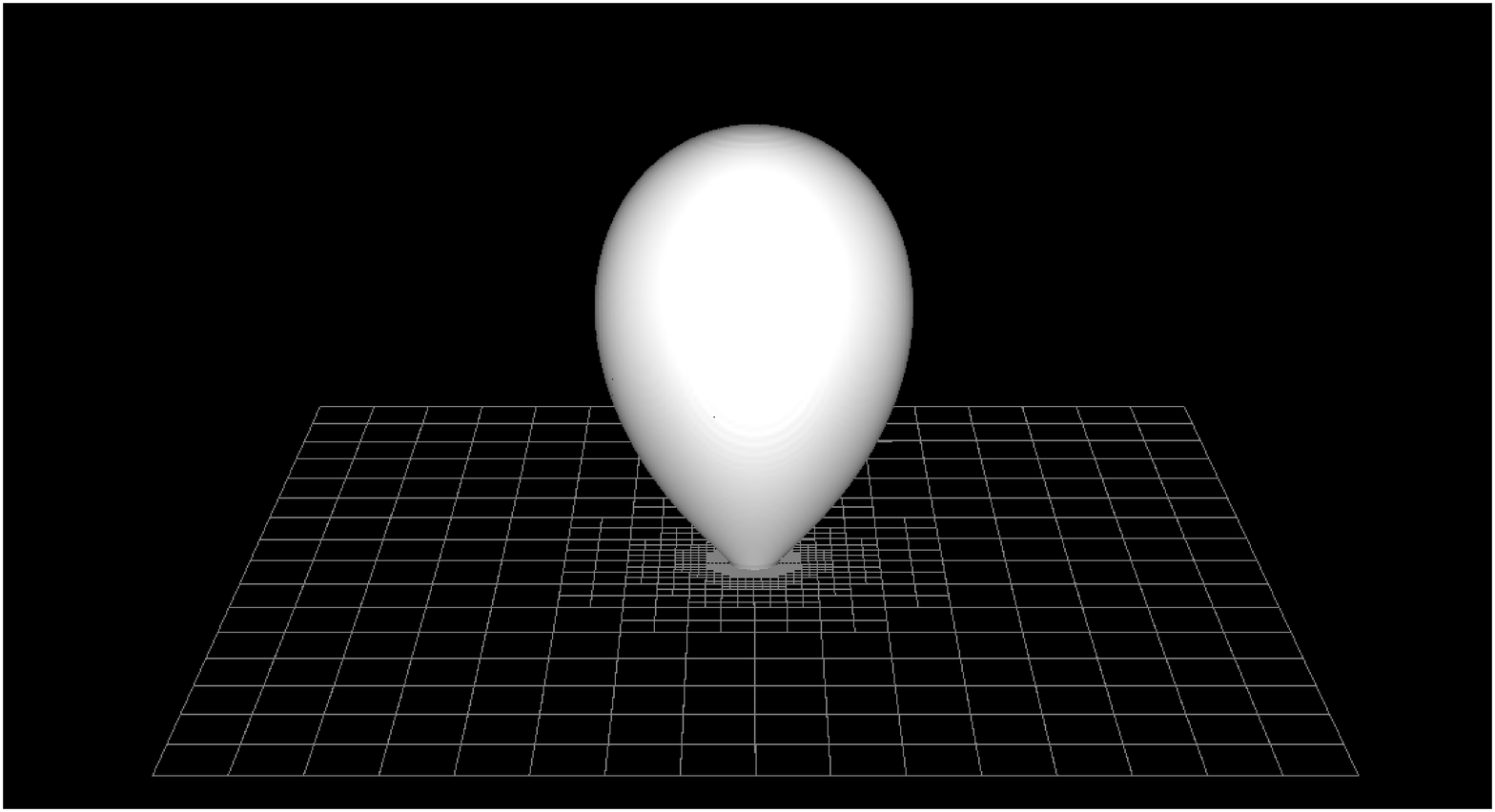}&
  \includegraphics[width=0.235\textwidth]{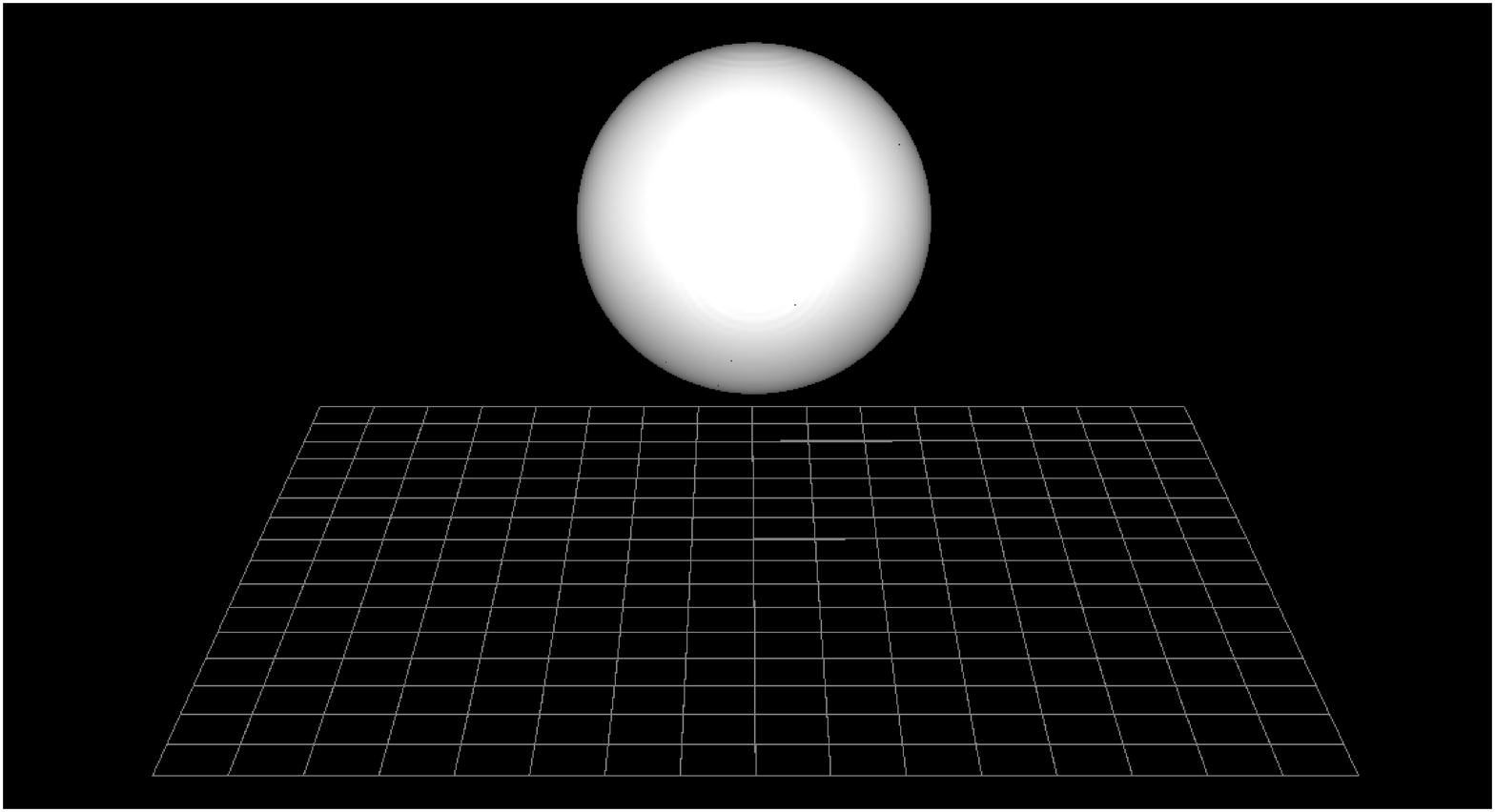}\\  
100 ps&190 ps  \\
&\hspace{-47mm}(a)\\
\includegraphics[width=0.25\textwidth, trim=4mm 2mm 0 0]{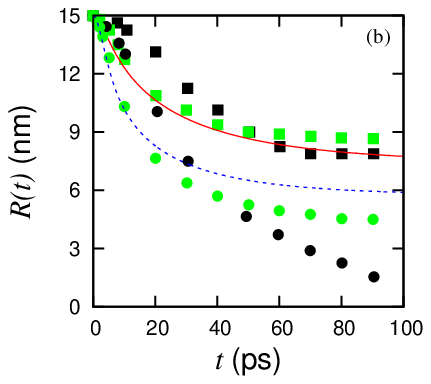} &
\includegraphics[width=0.25\textwidth, trim=4mm 2mm 0 0]{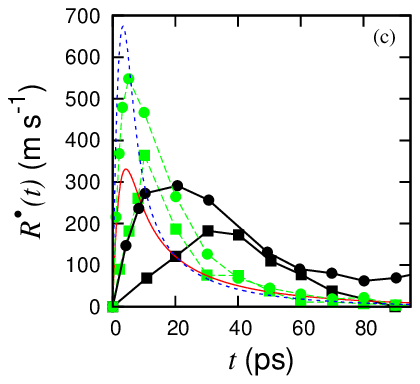}
\end{tabular}
\caption{
(Color online) Evolution of a disk ($h_0=15$ \AA, $R_0=150$ \AA, $\theta_{0}=90^\circ$); 
(a) for $\theta_{\text{eq}}=140{^\circ}$, the disk detaches at around $100$ ps (see also the animation in~\cite{sup_mat});
(b) $R(t)$ and (c) $\dot R(t)$ for  $\theta_{\text{eq}}=80{^\circ}$ ({\small $\blacksquare$})/({\color{green}{$\blacksquare$}})
and $140{^\circ}$ ($\bullet$)/({\color{green}{$\bullet$}})
when $Oh\approx0.35$ (black)  and $Oh\approx1.1$ (green).  Predictions of the model, 
Eq.~(\ref{eq:energy}), for $\theta_{\text{eq}}=80^\circ$ (solid red line) and 
$\theta_{\text{eq}}=140^\circ$ (dashed blue line) when  $Oh\approx1.1$.}
\vspace{-0.1in}\label{fig:3D140}
\end{figure}

{\it Results.} Figure~\ref{fig:3D80}(a)  shows the evolution of a Cu disk 
of initial height $h_0=15$ \AA~and radius $R_0=150$ \AA~with an initial
contact angle $\theta_{0}=90^\circ$, when the equilibrium contact angle $\theta_{\text{eq}}=80^\circ$,
that dewetts and collapses into a spherical cap.  Figure~\ref{fig:3D80}(b) and (c) 
show the front position, $R(t)$, and velocity, $\dot R(t)$, respectively.
We also show the results of the MD simulations obtained by employing two 
different Lennard-Jones (LJ) potentials,  LJ-a and LJ-b, that differ by the depth
of potential well \cite{fuentes_pre11}.
 These results show that
our numerical results and the MD simulations yield fully consistent retraction time scales. 
We also see that $\theta_{\text{eq}}$ influences the initial rapid retraction, with larger $\theta_{\text{eq}}$ 
leading to faster dynamics for early times.

A different type of evolution occurs if $\theta_{\text{eq}}$ is further increased.
To illustrate this, Fig.~\ref{fig:3D140}(a) shows snapshots of the profiles for
$h_0=15$ \AA, $R_0=150$ \AA, $\theta_{0}=90^\circ$, and $\theta_{\text{eq}}=140^\circ$.   For this
$\theta_{\text{eq}}$, the contraction is so fast that the nanodrop jumps off the 
surface,  following elongation in the $y$ direction (normal to the substrate).   Similar behavior is 
observed for any $\theta_{\text{eq}} \agt 130^\circ$, again in quantitative agreement
with the MD simulations~\cite{fuentes_pre11}.

To gain additional insight regarding the effects that drive contraction, and 
possibly ejection of the fluid, we have carried out additional set of 
simulations with modified inertial effects.  This was implemented
by varying incompressible fluid density, but more generally the 
dynamics can be characterized by Ohnesorge number, $Oh=\eta/\sqrt{\rho\sigma R_0}$,
which is related to  the Reynolds number defined based on a characteristic 
capillary velocity,  $\sigma/\eta$, as $Re=Oh^{-2}$. For liquid Cu results in Fig.~\ref{fig:3D80}, 
$Oh\approx0.35$, suggesting that inertial effects are important; the same conclusion can 
be reached by considering an intrinsic length scale,  $\ell_v=\eta^2/(\rho\sigma)$, above which inertial effects 
become significant; for liquid Cu, $\ell_v\approx 1.78$ nm, therefore smaller than typical length scales considered here.
As an example, Fig. \ref{fig:3D140}(b) and (c) show $R(t)$ and $\dot R(t)$ for
$Oh\approx0.35$ and $Oh\approx1.1$.  Reduced inertial effects eliminate the most noticeable feature of the dewetting:
nanodroplets do not detach from the surface for the same $\theta_{\text{eq}}$. Additional simulations
have shown that even for $\theta_{\text{eq}}=150^{\circ}$ there is no detachment
for $Oh\approx1.1$. 

A basic understanding of the effects that drive the fluid evolution can be reached by a relatively simple effective
model based on the balance of relevant energies. In this model, the evolution of the nanodisk obeys the energy
balance equation
\begin{equation}
 \frac{\partial}{\partial t}\left[E_k+E_s\right]+D=0~,\label{eq:energy}
\end{equation}
where $E_k$, $E_s$ are the kinetic and surface energy, respectively, and $D$
is the rate of energy loss due to viscous dissipation, neglecting the gravitational energy.
In this model, we describe the nanodisk dynamics by a fluid cylinder evolving on a solid substrate;
a similar model has been considered for a drop impact
problem~\cite{model}. Here, 
$E_k = {\rho \int_{\Omega}|\vec{V}|^2d\,\Omega/2}$
where the integration  is over the fluid cylinder, and $\vec V=\vec V(\vec x,t)$ is the axially symmetric fluid velocity.   
Using Young's law \cite{Young}, $E_s = \sigma\left[\pi R(t)^2(1-\cos\theta_{\text{eq}})+2\pi R(t)h(t)\right]$, 
where $R(t)$ is the wetted radius and $h(t)$ is the cylinder height. The viscous dissipation energy
$D=\int_{\Omega}(\tilde{\tau}: \nabla \vec{V}) d\,\Omega$, where $\tilde{\tau}$ is the shear stress 
tensor. Using axisymmetric stagnation point flow and the conservation of mass, Eq.~(\ref{eq:energy}) reduces to a nonlinear second-order
variable-coefficient ordinary differential equation given in~\cite{sup_mat}. This equation is then solved numerically, and the results
are shown together with the computational results in Figs.~\ref{fig:3D80} and \ref{fig:3D140} (plane lines).
We find very reasonable degree of agreement, suggesting that this simple model captures well
the main mechanisms driving fluid contraction.

\begin{figure}[t]
\centering
\begin{tabular}{cc}
\includegraphics[width=0.2475\textwidth, trim=2mm 2mm 0 0]{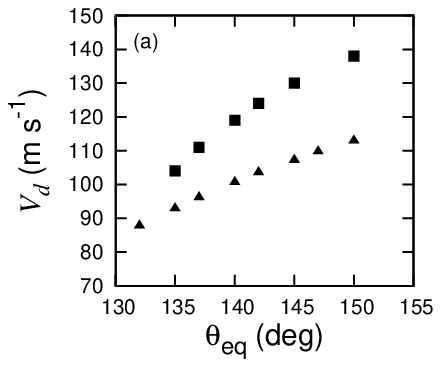} &
\includegraphics[width=0.24\textwidth, trim=4mm 2mm 0 0]{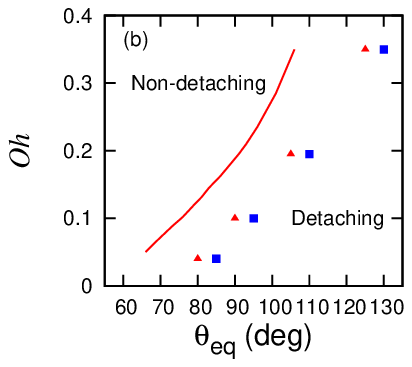} 
\end{tabular}
\caption{(Color online)
(a) The effect of $\theta_\text{eq}$
on the ejection velocity, $V_d$, for a Cu nanodisk
($h_0=15$ \AA, $R_0=150$ \AA) predicted by the model based on Eq.~(\ref{eq:energy}) ($\blacktriangle$), see
also Eq.~(13) in~\cite{sup_mat},  
and simulations ({\small $\blacksquare$}). (b)
Phase diagram showing the influence of $Oh$ and $\theta_\text{eq}$ on 
the ejection.  The solid line is based on the model,  Eq.~(\ref{eq:energy}), see  
also Eq.~(13) in~\cite{sup_mat}, and the symbols mark the result of the numerical simulation 
showing detaching ({\color{blue}{$\blacksquare$}}) and non-detaching 
({\color{red}{$\blacktriangle$}}) nanodrops. A linear fit to the solid line yields a slope of $0.0075$.}
\label{fig:model}
\end{figure}

Figure~\ref{fig:model} shows additional comparison between the numerical simulation and  the effective model.   
In Fig.~\ref{fig:model}(a), we observe that both the simulations and the model, Eq.~(\ref{eq:energy}), 
predict approximately linear dependence of the ejection velocity on $\cos \theta_{\text{eq}}$.
We expect that the discrepancy between the two is maily due to the deficiency of the model for 
large $\theta_\text{eq}$~\cite{sup_mat}. 
Figure~\ref{fig:model}(b) show a phase diagram in $(Oh,~\theta_{\text{eq}})$ parameter space illustrating
the criteria for ejection resulting from the model, together  with the  results of the numerical simulation, concentrating here 
on small values of $Oh$.
Although the model underpredicts $\theta_\text{eq}$ required for ejection, we note
a consistent trend of the results.   Note that for large $Oh\approx 1.1$ considered previously, both the model 
and the simulation are in non-detaching regime. Further investigation of the  results of the model
suggests that the ejection velocity increases  approximately linearly as $Oh$ decreases
when $\theta_\text{eq}$ is fixed (see \cite{sup_mat} for more details). 

\begin{figure}[t]
\centering
\begin{tabular}{cc}
  \includegraphics[width=0.235\textwidth]{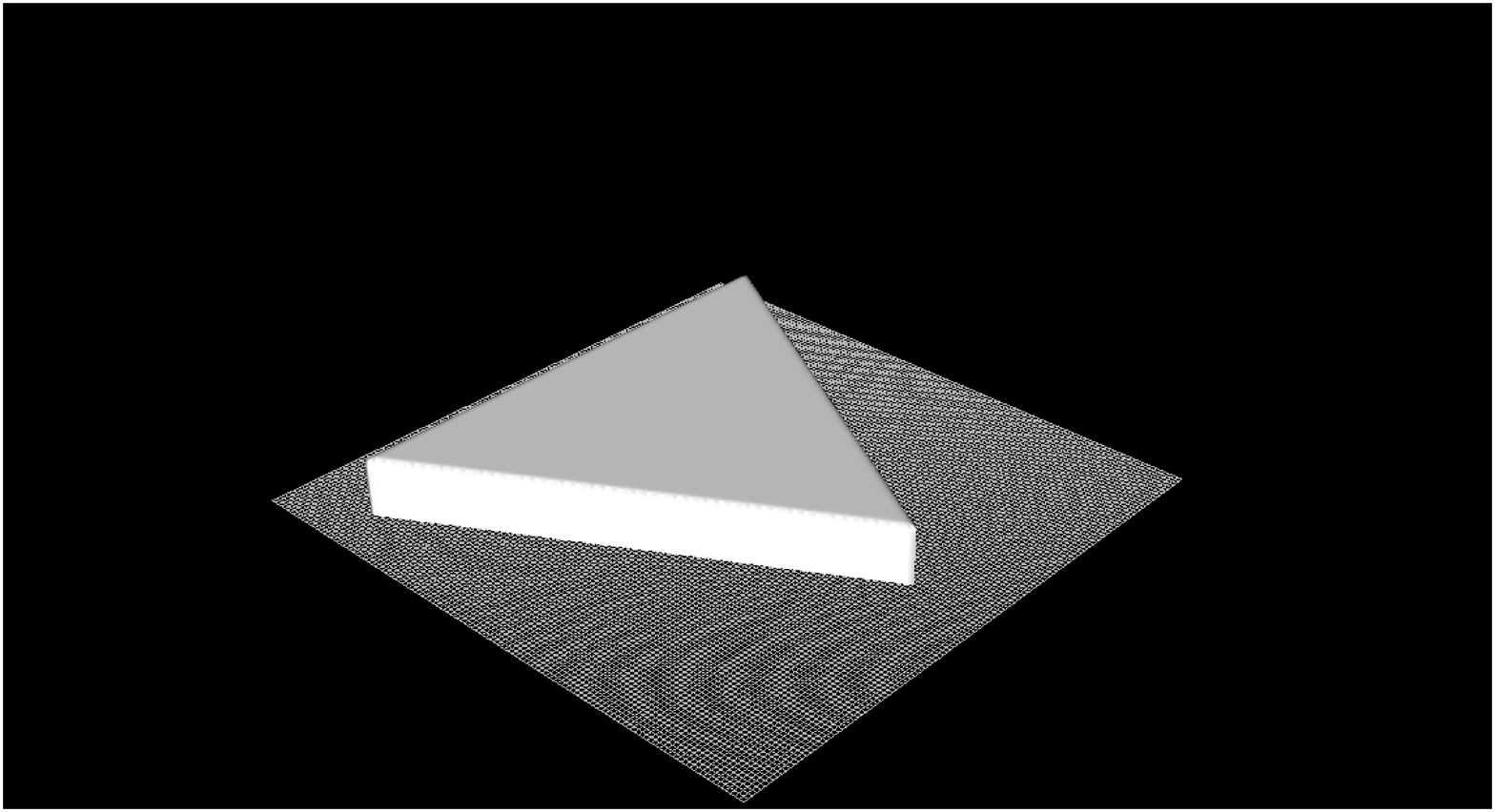}&
  \includegraphics[width=0.235\textwidth]{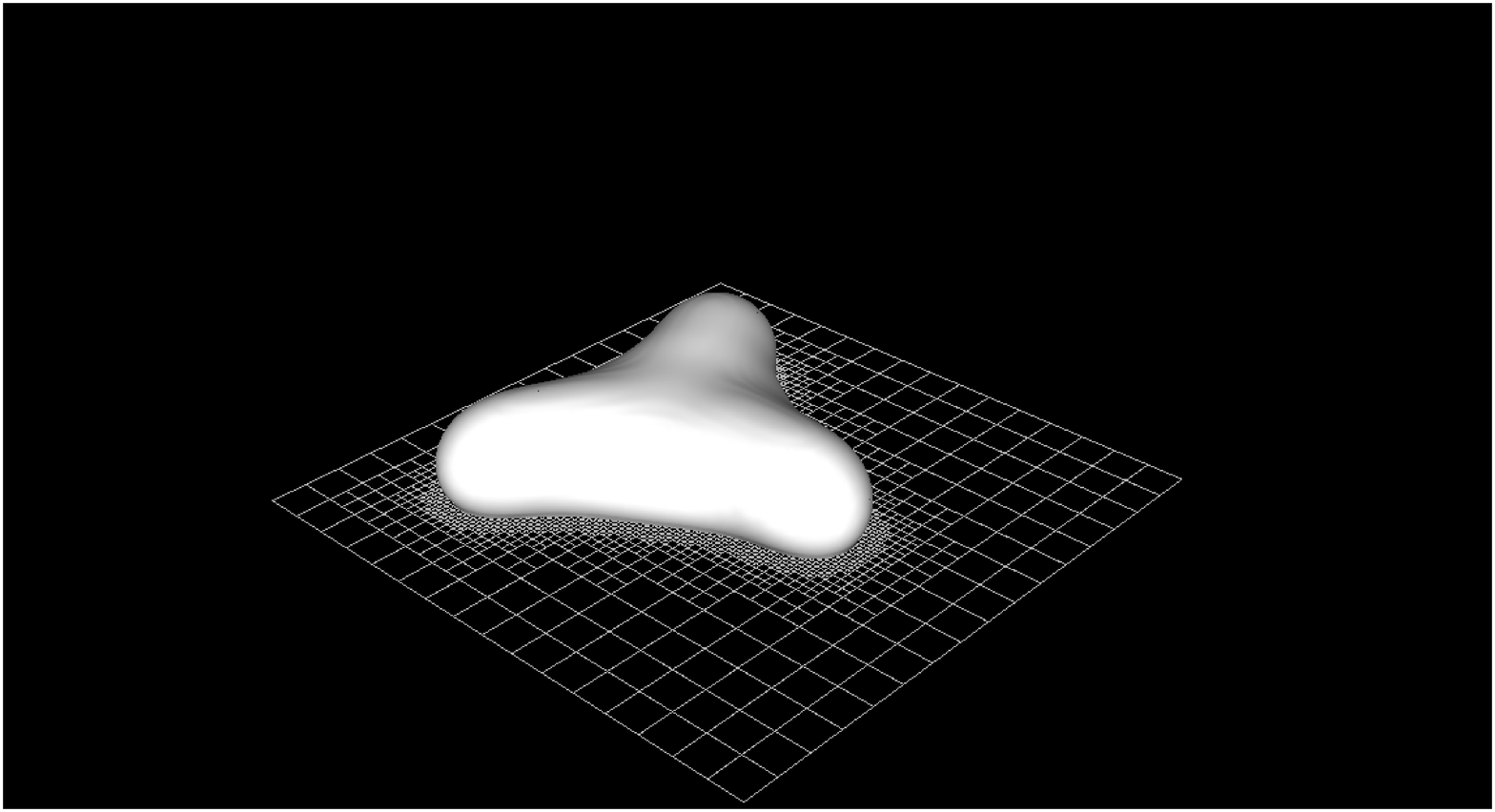}\\  
  $0$ ns& $1.5$ ns \\
  \includegraphics[width=0.235\textwidth]{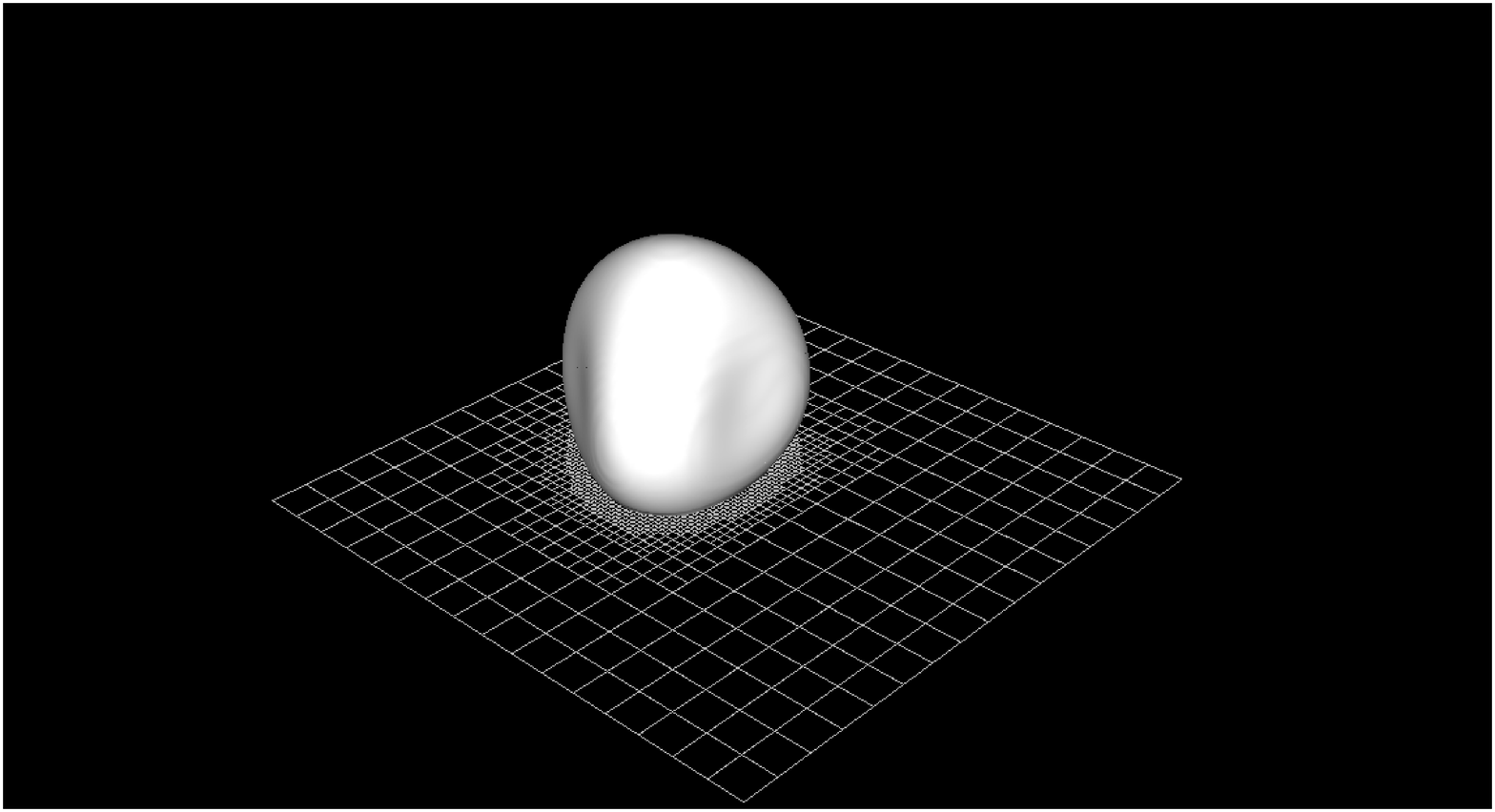}&
  \includegraphics[width=0.235\textwidth]{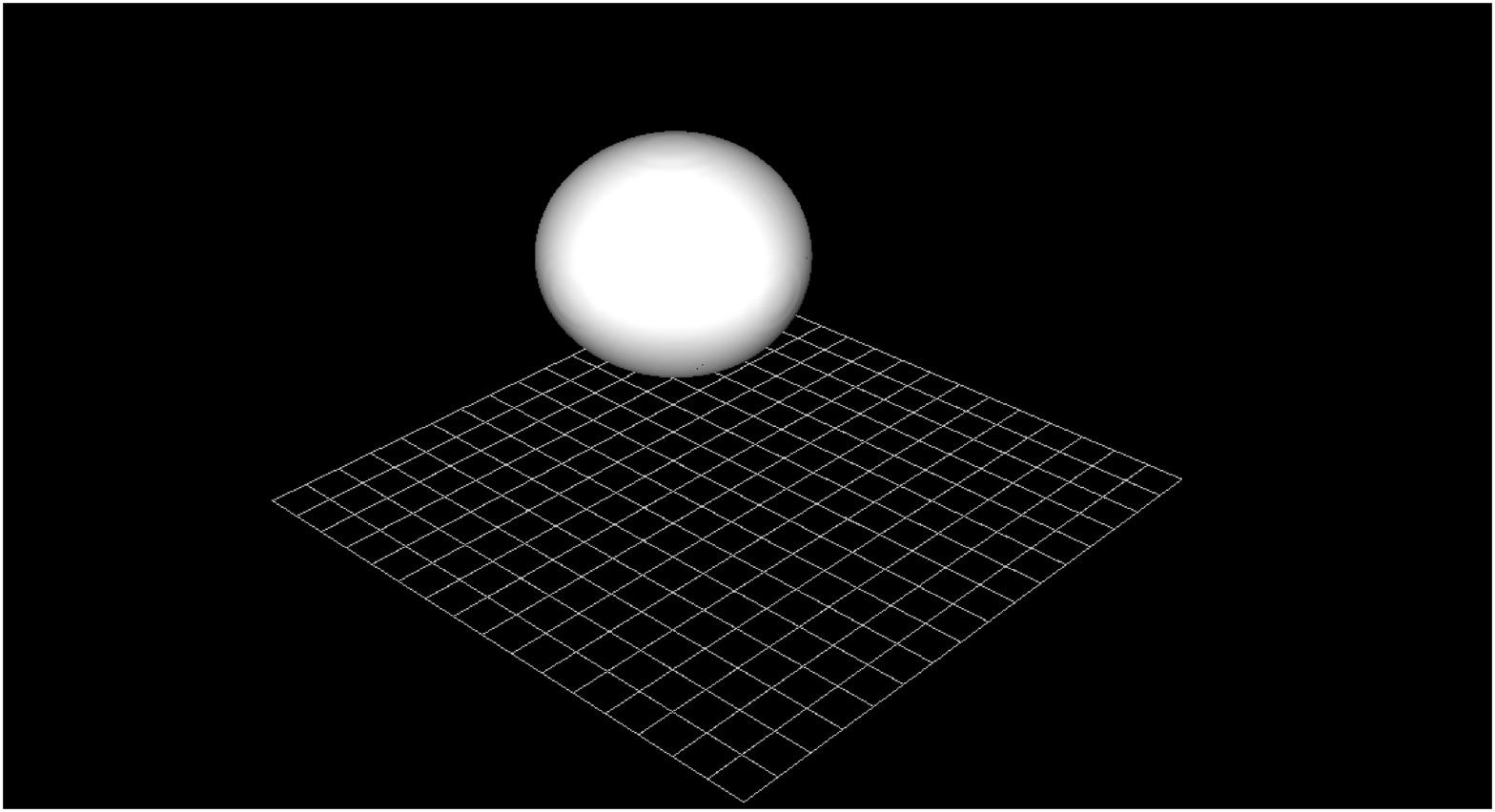}\\ 
  $4$ ns&   $10$ ns
\end{tabular}
\caption{
Evolution of a Au equilateral triangle ($a = 405$ nm, $h_0 = 47$ nm,
$\theta_{0}=90^\circ$, and $\theta_{\text{eq}}=140^\circ$). The structure
collapses into a droplet and detaches at $t\approx 7$ ns  (see also the animation in~\cite{sup_mat}).}
\vspace{-0.1in}\label{fig:H0.1}
\end{figure}
\begin{figure}[tbh]
\centering
\begin{tabular}{cc}
  \includegraphics[width=0.235\textwidth]{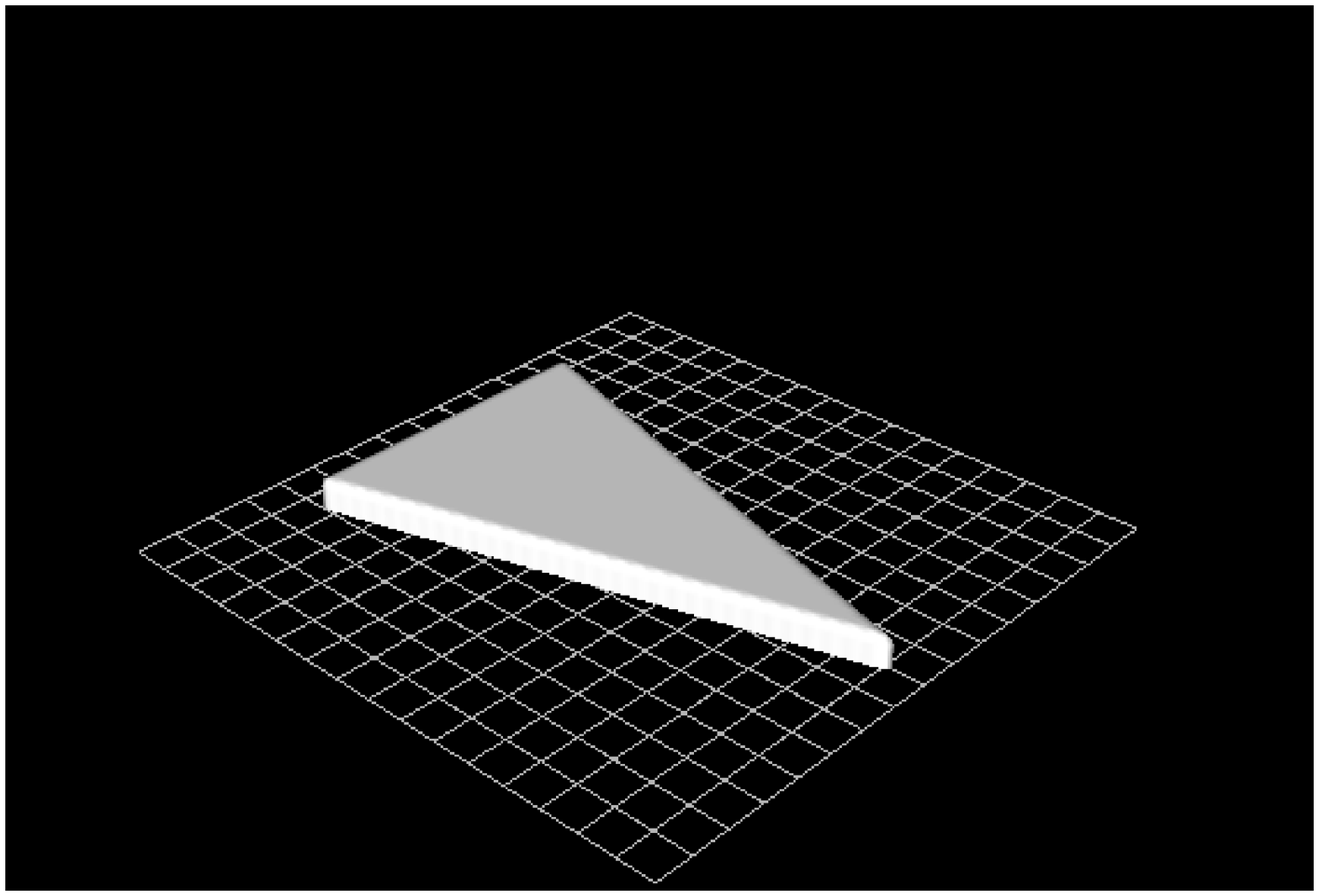}&
  \includegraphics[width=0.235\textwidth]{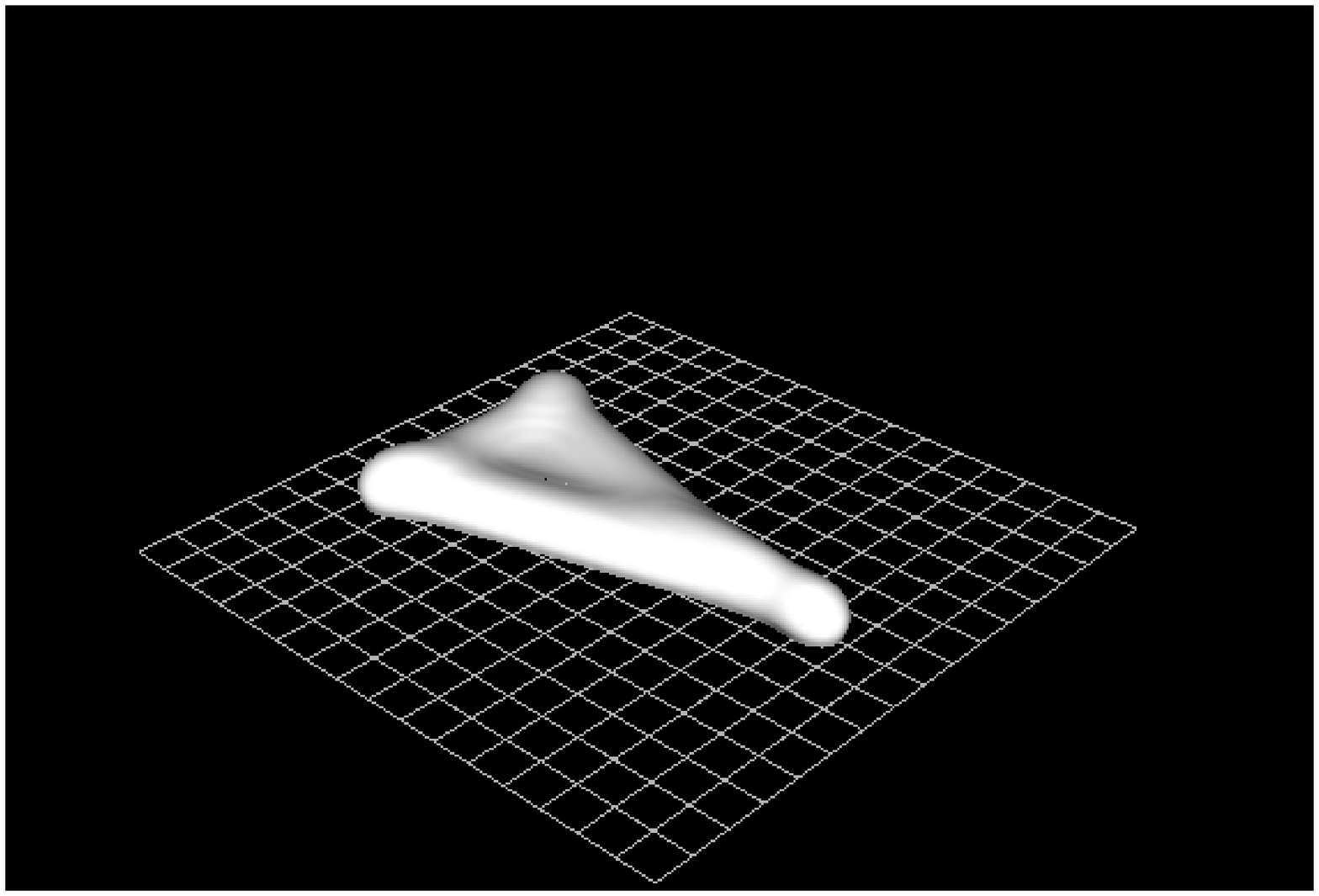}\\
   0 ns &  0.6 ns \\  
  \includegraphics[width=0.235\textwidth]{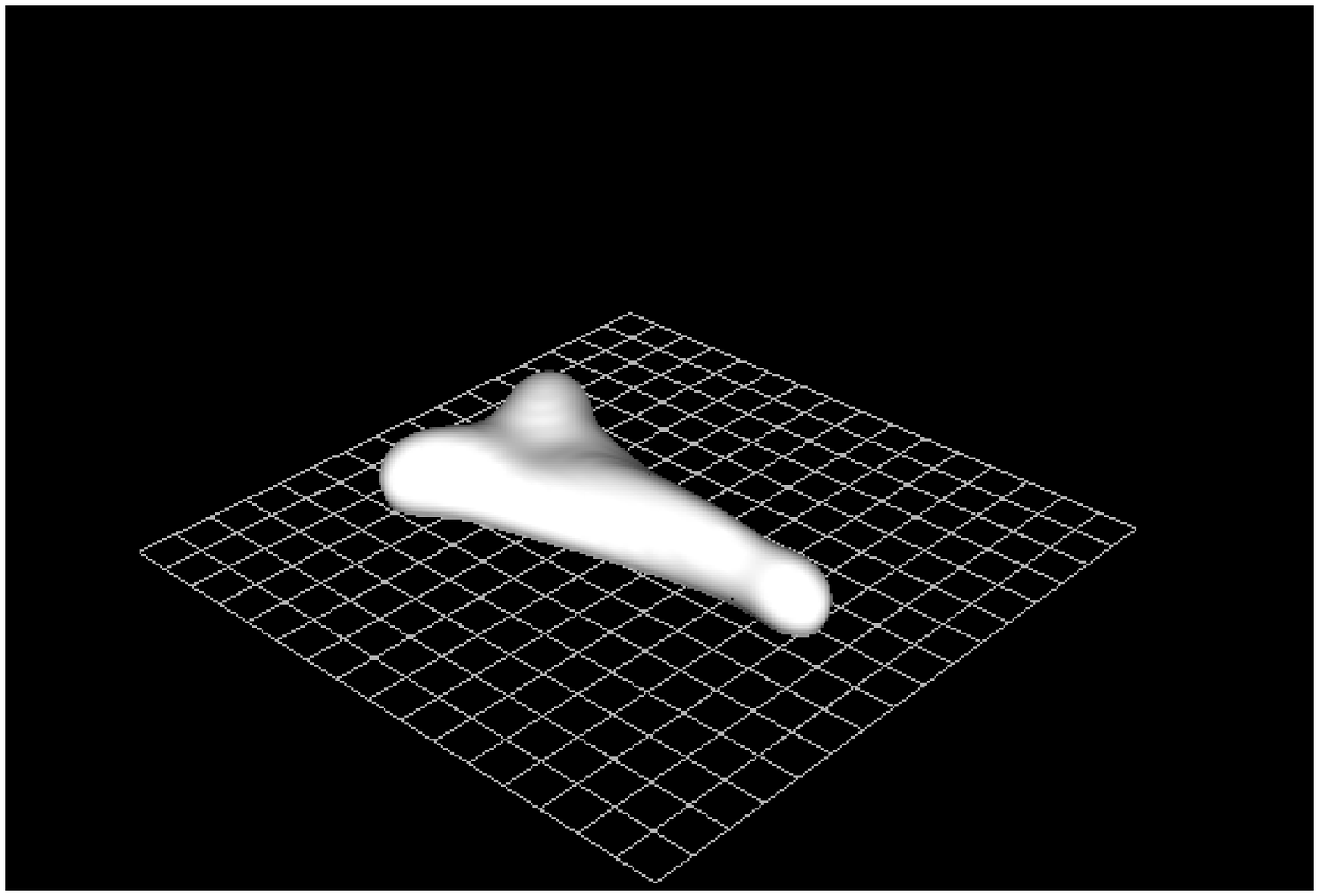}  &
  \includegraphics[width=0.235\textwidth]{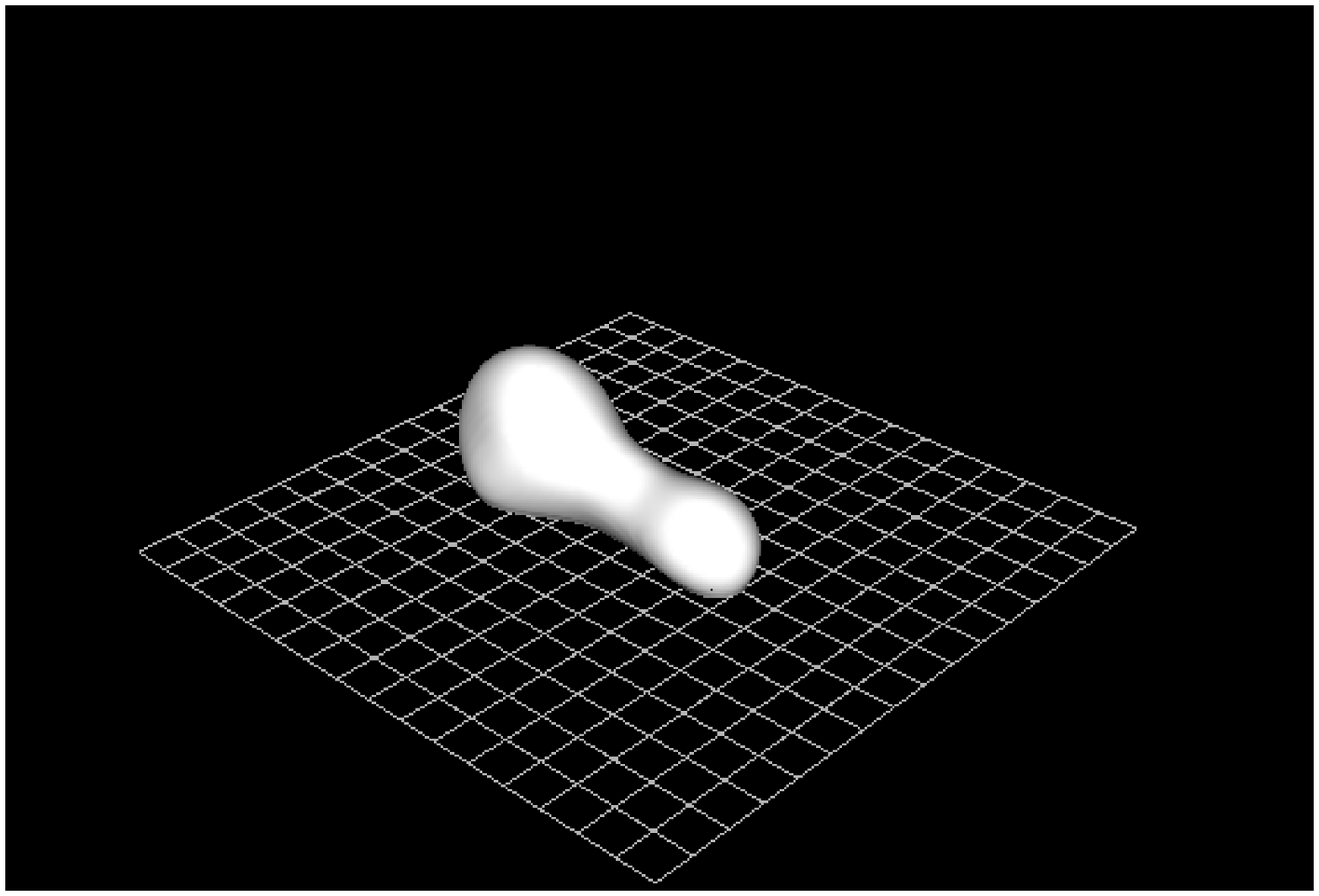}\\
   0.9 ns &  2 ns \\  
  \includegraphics[width=0.235\textwidth]{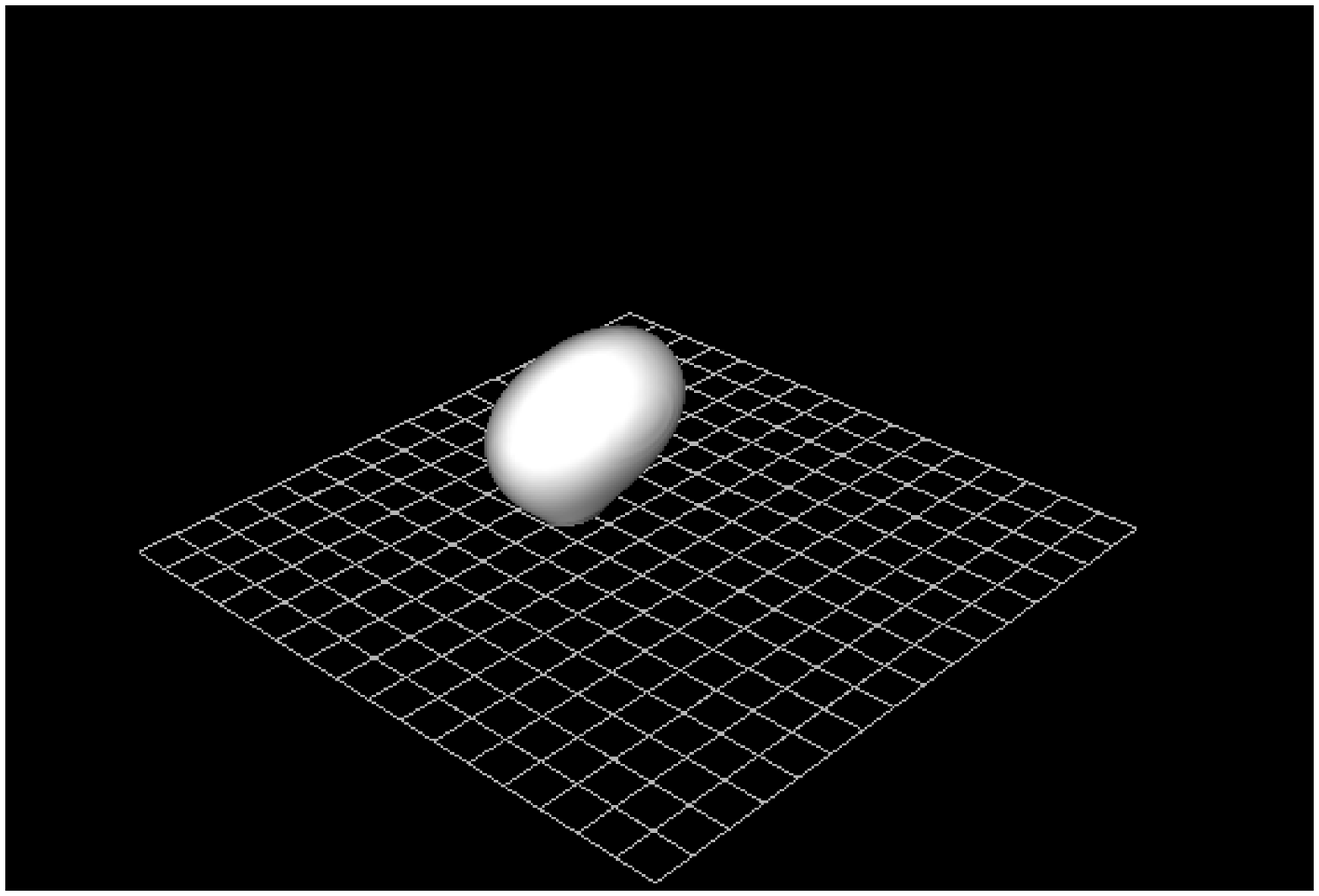}&
  \includegraphics[width=0.235\textwidth]{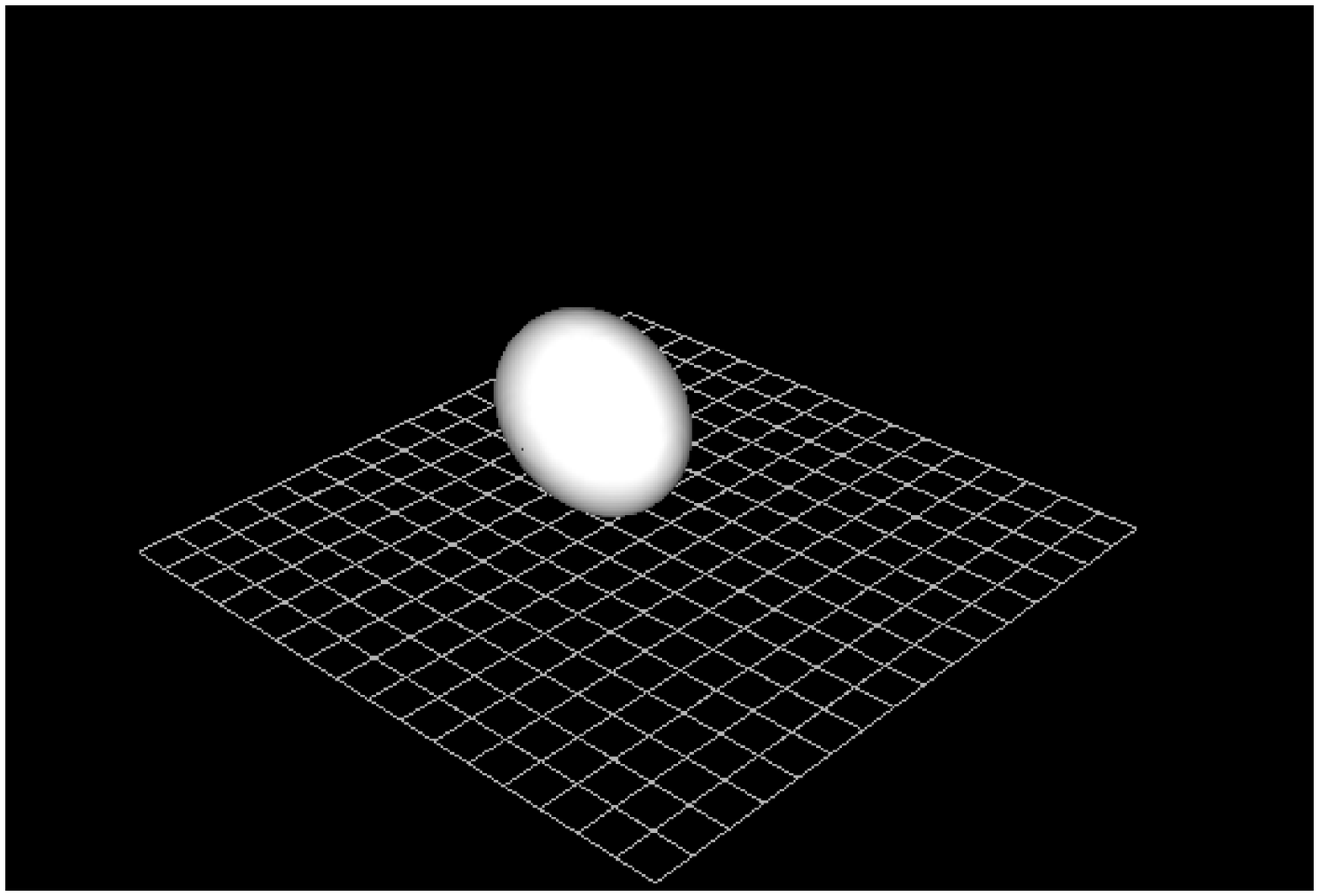}\\
   5 ns & 7.4 ns   
\end{tabular}
\caption{
Evolution of a Au isosceles triangle for $h_0= 24$ nm, $\theta_{0}=90^\circ$, 
long sides $a = 438$ nm and short side $b= 247$ nm;
$\theta_{\text{eq}}=140^\circ$. The triangle
collapses into a droplet and detaches from the substrate at around 5 ns (see also the animation in~\cite{sup_mat}).
}
\vspace{-0.1in}\label{fig:H0.05-2x1}
\end{figure}
Next we consider the configuration representative of physical experiments where Au triangular structures
were liquefied and let to evolve on SiO$_2$ substrate~\cite{habenicht_05}. 
Figure \ref{fig:H0.1} shows the snapshots of initially equilateral triangle ($\theta_{0}=90^\circ$).
The dewetting process first starts at the vertices, where
the curvature is high. Due to high surface tension and large
$\theta_\text{eq}$, the fluid starts to accumulate there.
The humps at the vertices then coalesce into a droplet.
Owing to a low viscosity of liquid gold, inertial effects dominate over
viscous dissipation (here $Oh \approx 0.047$ \cite{sup_mat}), 
giving rise to an upward movement that leads
to droplet detaching from the surface with a velocity of $\approx 24$ m s$^{-1}$.
This process is consistent with the dewetting induced ejection
mechanism outlined in \cite{habenicht_05}. We note the consistency 
of the time scales found by the simulations and the experiments in \cite{habenicht_05},
where an ejection time scale of the order of $10$ ns was observed. In addition to comparing
favorably with experiments, the simulations allow for
an additional insight since it is possible to explore very fast
time scales, and also easily vary the parameters,
such as $\theta_\text{eq}$ and the initial shape, and ask what is their influence
on the outcome.
\begin{figure}[tbh]
\centering
\begin{tabular}{cc}
  \includegraphics[scale=0.45, trim=10mm 15mm 0 0, clip=true]{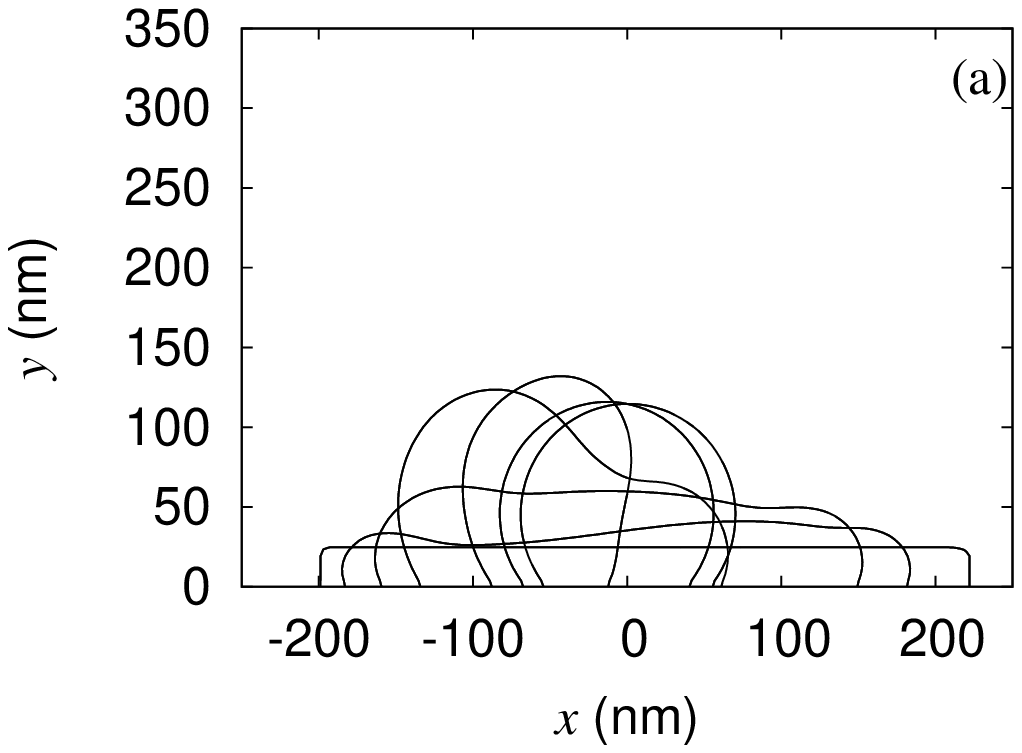}&
  \includegraphics[scale=0.45, trim=32mm 15mm 0 0, clip=true]{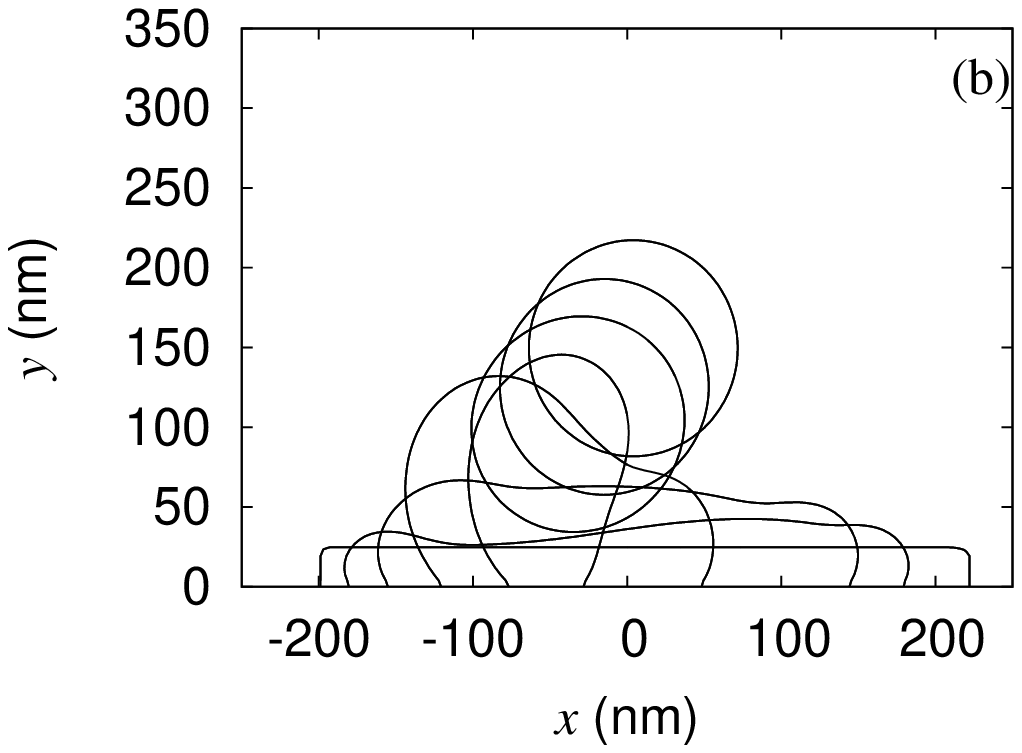}\\  
  \includegraphics[scale=0.45, trim=10mm 0 0 0, clip=true]{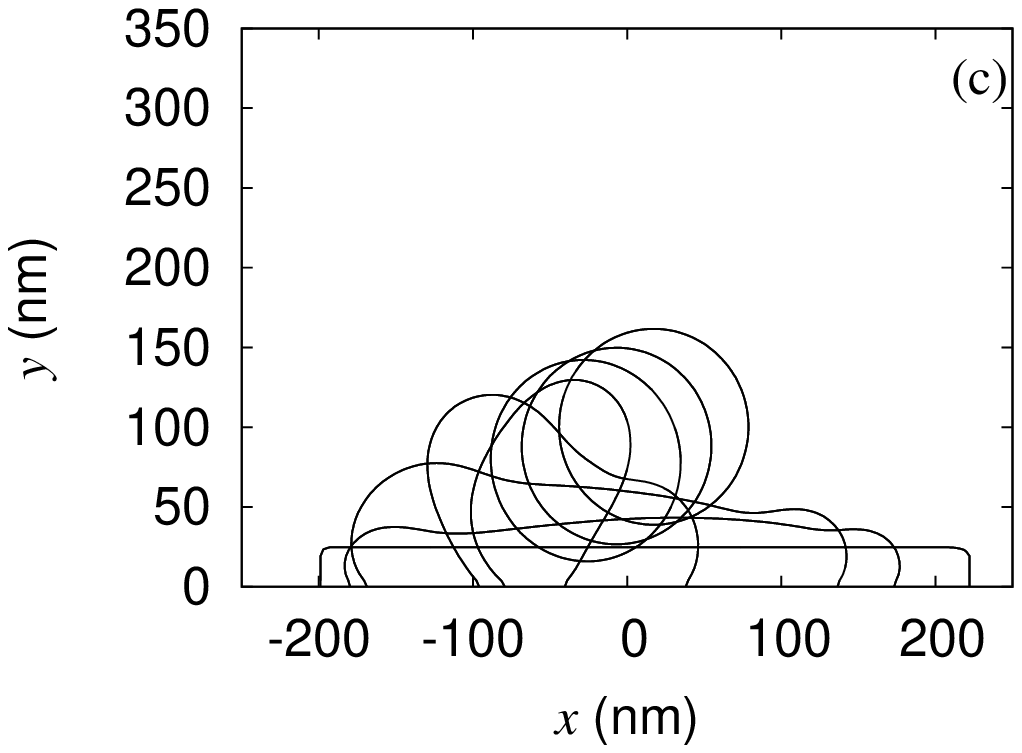}&
  \includegraphics[scale=0.45, trim=32mm 0 0 0, clip=true]{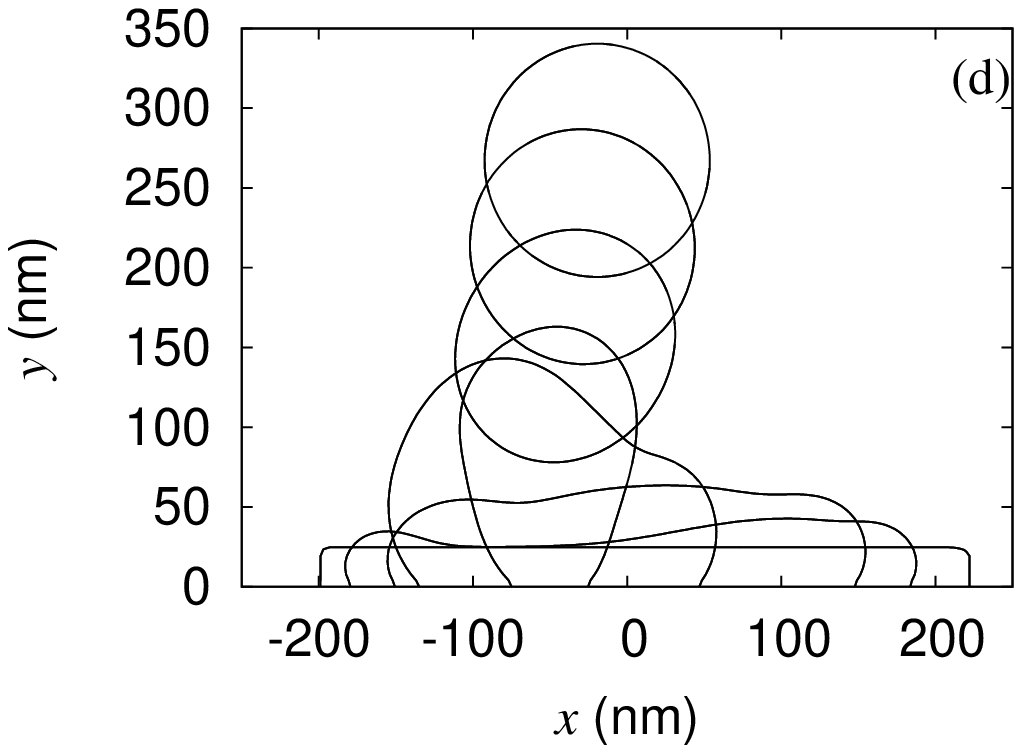}\\
 \includegraphics[width=0.25\textwidth, trim=4mm 0mm 0 0]{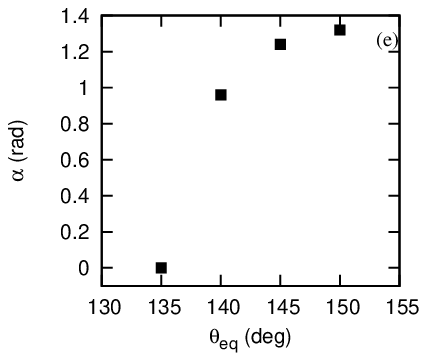}&
 \includegraphics[width=0.25\textwidth, trim=4mm 0mm 0 0]{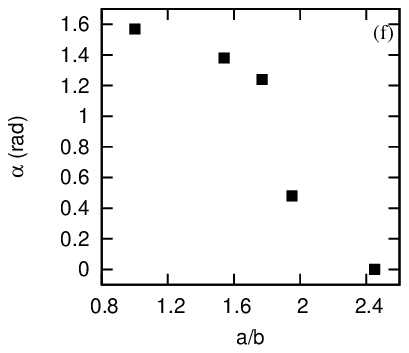} \\  
\end{tabular}
\caption{$x$-$y$ cross sections of Au isosceles triangles
for $h_0= 24$ nm and $\theta_{0}=90^\circ$; (a-b) $a = 438$ nm and $b=247$ nm
for (a) $\theta_{\text{eq}}=130^\circ$, (b) $140^\circ$,
showing the increase of ejection angle, $\alpha$, as 
$\theta_{\text{eq}}$ is increased (note that in (a), $\alpha = 0$); 
(c-d) $\theta_{\text{eq}}=145^\circ$, and (c) $a = 362$ nm, $b=185$ nm, 
(d) $a =475$ nm, $b=308$ nm; 
$t=0,~ 0.6,~ 1.2,~ 3,~ 4.4,~ 11,~ 19,~ 35$ ns.
(e-f) Ejection angle $\alpha$ versus the $\theta_{eq}$ (e), when $a/b\approx1.77$,
and the initial geometrical ratio $a/b$ (f), when $\theta_{\text{eq}} = 145^\circ$.
}
\label{fig:H0.05-2x1-theta}
\end{figure}

To illustrate the influence of geometry, we consider isosceles triangles.  
Figure \ref{fig:H0.05-2x1} shows a perhaps unexpected result, that
despite the fact that the vertex at the smallest angle collapses the fastest
due to a higher curvature, the
vertices at the larger angles arrive sooner to the center (future work should
analyze how general this result is).  This mismatch 
excites oscillations of the droplet translating into a tumbling movement 
after the ejection. 

Next, we ask whether the ejection angle, $\alpha$, is always $\pi/2$.   We find that $\alpha$ can be influenced 
for asymmetric collapse, and present two means for controlling it by modifying:  (i) $\theta_{\text{eq}}$ and (ii) the initial geometry.   
Figure \ref{fig:H0.05-2x1-theta}(a-b) shows that
increasing $\theta_{\text{eq}}$ alters the no-ejection ($\alpha = 0$) to a directional ejection ($0\le \alpha \le {\pi/2}$).
A qualitative understanding can be reached by recalling that 
$\alpha \ne {\pi/2}$ is due to the lack of synhronicity of the collapse.   
A larger $\theta_{\text{eq}}$ results in higher surface energy, speeding up the fluid at the vertices.
Therefore for larger $\theta_{\text{eq}}$, collapse is more synchronous and $\alpha$ is closer to $\pi/2$.
For smaller $\theta_{\text{eq}}$, $\alpha$ may vanish  (see Figure \ref{fig:H0.05-2x1-theta}(a)), \ie ejection does not 
happen, while an equilateral triangle with the same $\theta_\text{eq}$ does eject, illustrating the effect
of asymmetry.  We next consider the aspect ratio of the lengths, ($a,b$), of triangle sides,
and find that the larger $a/b$ ratios lead 
to increased asymmetry of the collapse and smaller $\alpha$, as shown in Fig.~\ref{fig:H0.05-2x1-theta}(c-d).

Figures \ref{fig:H0.05-2x1-theta}(e) and (f) show $\alpha$ versus $\theta_\text{eq}$ for fixed $a/b$, and
versus $a/b$ for fixed $\theta_\text{eq}$, respectively.
In Fig. \ref{fig:H0.05-2x1-theta}(e) we see that the ejection angle
increases monotonically with an increase of $\theta_\text{eq}$ with no ejection
for $\theta_\text{eq} \alt 135^\circ$. Figure \ref{fig:H0.05-2x1-theta}(f) shows that 
$\alpha$ decreases with increased asymmetry and that  
for the considered $\theta_\text{eq}$, the ejection does not occur, \ie $\alpha=0$, when $a/b\agt2.4$.
Very large ratios lead to a different type of dynamics: Rayleigh-Plateau type
of breakup, see~\cite{Kondic} for discussion in the context of liquid metals.  
More elaborate analysis of the influence of geometry and wetting properties is needed to fully 
understand the dynamics - we leave this for future work.

{\it Conclusions.}  
In this work, we have demonstrated that  continuum simulations provide a
good qualitative agreement with both MD simulations on the length scales in the
range of $1$-$10$ nm, and with the physical experiments with in-plane length scales measured
in the range of $100$ nanometers.    
We expect that this finding will further motivate modeling and computational
work on these scales, since it is suggesting that the continuum-based simulation has a
predictive power. For the problem of dewetting and possibly detaching nanodrops,
the simulations provide precise insight regarding the influence of 
inertial, viscous, and capillary forces, in addition to the liquid/solid interaction.
This insight is also confirmed by a simple model based on an energy balance and accounting
for viscous energy losses. Furthermore, the simulations provide a clear prediction that the
direction of ejection of fluid from the substrate can be influenced (and controlled)
by modifying either the wetting properties or the initial geometry. Future computational
and modeling work will include the thermal and phase change effects, as well
as more accurate modeling of liquid-solid interaction in the vicinity of contact lines,
allowing to  even more accurately model the dynamics of liquid metals on nanoscale.

\clearpage
{\bf Supplementary Material: Numerical simulation of ejected molten metal-nanoparticles liquefied
       by laser irradiation: Interplay of geometry and dewetting}\\

\section {Mathematical model}   
The equations of conservation of mass, $\nabla \cdot \vec{U} = 0$, and momentum are
written as
\begin{equation}
  \frac{\partial}{\partial t} (\rho \vec{U}) + 
  \nabla \cdot (\rho \vec{U}\vec{U}) = -\nabla p
  + \nabla \cdot \tilde{\tau} 
  + \vec{F}_{st}\, , 
  \label{eq:momentum}
\end{equation}
where $\vec{U}=(u,v,w)$ represents the velocity vector, $p$ the pressure, $\rho$ the fluid density, $\tilde{\tau}$
the shear stress tensor, and $\vec{F}_{st}$ the surface tension force (per unit volume) acting on the fluid.
The shear stress 
tensor is defined as $\tilde{\tau}=\eta(\nabla \vec{U}+(\nabla \vec{U})^{T})$, where $\eta$ represents the
fluid dynamic viscosity. 

The flow equations have been written in an Eulerian frame of reference, and thus a solution of these equations
must be coupled with a methodology for following the deforming fluid-fluid interface. Here, the `Volume of Fluid'
(VoF) algorithm is implemented \cite{Hirt81,Gueyffier98,AB2009}. Volume tracking requires the introduction of a scalar 
function $f$ defined as
\begin{eqnarray*}
  f & = & \left\{ \begin{array}{ll}
	0 & \mbox{in fluid 1} \\
	1 & \mbox{in fluid 2}, \end{array} \right.
\end{eqnarray*}
for a two fluid system. Since $f$ is passively advected with the flow, it satisfies the advection equation
\begin{equation}
  \frac{\partial f}{\partial t} + (\vec{U} \cdot \nabla) f = 0.
  \label{eq:vofadv}
\end{equation}
Density and viscosity are then evaluated via volume-weighted formulae as $\rho = \rho_1 + (\rho_2 - \rho_1) f$
and $\eta = \eta_1 + (\eta_2 - \eta_1) f$, respectively, where subscripts 1 and 2 refer to fluids 1 and 2, respectively.

The surface tension force $\vec{F}_{st}$ in (\ref{eq:momentum}) is reformulated as an equivalent
volume force \cite{Brackbill92}, which 
is non-zero only at each interface cell
\begin{equation}
\vec{F}_{st} = \sigma \kappa \delta_S \hat{n},
\end{equation}
where $\sigma$ is a constant interfacial  tension
and $\delta_S$ denotes the Dirac delta function for the surface separating the
fluids.
The curvature $\kappa$ and unit normal $\hat{n}$ directed into fluid 1 
are geometric characteristics of the surface and are described in terms of $f$ and
computed with a second-order `height-function' (HF) method \cite{Suss2003,Cummins2005,AB2008,FCDKSW2006}.
Within the VoF-based sharp surface tension
representation,
$\delta_S \hat{n}$ is equivalent to $\nabla f$, thus
\begin{equation}
\vec{F}_{st}  = \sigma \kappa \nabla f.
\label{eq:st}
\end{equation}

If partial slip is allowed at the contact line, we impose the Navier slip 
boundary condition \cite{Haley91-sup} at the (static) solid surface $y=0$, 
\begin{equation}
(u,w){|}_{y=0}=\lambda {\partial (u,w)/\partial y}{|}_{y=0},
\end{equation}
where $\lambda$ is the slip length.
For dynamic contact lines, the Navier slip boundary condition leads
to a regularization of the viscous stress singularity at the contact line \cite{Haley91-sup}.

\section {Computational model}  
The main issue is the imposition of the contact
angle, described in \cite{AB2008,AB2009}. Briefly,
within our VoF framework, the contact angle boundary condition 
enters the N-S solver in two ways: it defines the orientation 
$\hat{n}$ of the VoF interface reconstruction at the contact line,
and it influences the calculation of the contact line curvature;
together, these two reflect the wettability of the surface.
We implement the HF methodology to impose the contact angle 
in 3D \cite{AB2009}. The HF methodology accurately computes interface curvatures
at the contact line,
values that converge with mesh refinement.
In our HF methodology, the curvature at the contact line is computed  
at the desired contact angle in order to balance 
the surface tension forces on the contact line as in Young's relation \cite{Young-sup}
\begin{equation}
\sigma \cos \theta_{\text{eq}}=\sigma_{sg}-\sigma_{sl},\label{eq:Young} 
\end{equation}
where $\sigma_{sg}$ and $\sigma_{sl}$
are interfacial tensions for the solid-gas and solid-liquid surfaces, respectively. 
If the actual contact angle is out of equilibrium, the resulting mismatch in curvature
generates a kink in force thus driving the contact line to configuration that 
satisfies Young's relation. 

\section {Effective model}
\begin{figure}[h]
 \includegraphics[scale=0.5]{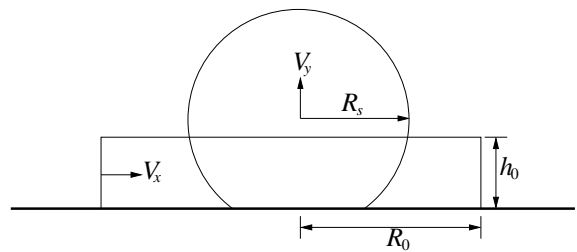}
 \caption{Cylindrical model of nanodisk retraction. $\vec V= (V_x,V_y)$ is the axially symmetric fluid velocity; 
          $x$ is the radial coordinate and $y$ is the height.}
\label{fig:1}
\end{figure}
Following \cite{kim2001,attane2007}, we developed a simple model for the evolution of the nanodisk based on the energy
balance. In this model, an ordinary differential
equation for the wetted radius $R(t)$ of a nanodisk  is developed 
by modeling the retraction dynamics of a fluid cylinder of radius $R$ and thickness
$h$ evolving on a solid substrate, see Fig.~\ref{fig:1}.
In this model, the kinetic energy of a drop of volume $\Omega$ is defined as
\begin{equation}
 E_k=\frac{\rho}{2}\int_{\Omega}(V_x^2+V_y^2)d\,\Omega.\label{eq:kinetic}
\end{equation}
The surface energy is defined as
\begin{equation}
 E_s=\sigma\left[\pi R(t)^2(1-\cos\theta_{\text{eq}})+2\pi R(t)h(t)\right].\label{eq:surface}
\end{equation}
The rate of energy loss due to viscous dissipation is defined as
\begin{eqnarray}
 D=2\eta\int_{\Omega}\left[\left(\frac{\partial V_x}{\partial x}\right)^2+ \nonumber
                          \left(\frac{V_x}{x}\right)^2 + \left(\frac{\partial V_y}{\partial y}\right)^2 \right. &+&\\ 
                          \left.
                          \frac{1}{2}\left(\frac{\partial V_x}{\partial y}+
                          \frac{\partial V_y}{\partial x}\right)^2\right]d\,\Omega.\label{eq:dissipation}
\end{eqnarray}
We assume an axisymmetric stagnation point flow
\begin{equation}
 V_y=\frac{-2}{R}\frac{dR(t)}{dt}y=\frac{1}{h(t)}\frac{dh(t)}{dt}y,\label{eq:vy}
\end{equation}
\begin{equation}
 V_x=\frac{1}{R}\frac{dR(t)}{dt}x=\frac{-1}{2h(t)}\frac{dh(t)}{dt}x.\label{eq:vx}
\end{equation}
We note that the velocity field assumed by Eqs.~(\ref{eq:vy}) and (\ref{eq:vx})
is shear free and so imply a slip at the solid substrate. This is consistent with the
assumption of a free-slip boundary condition specified at the substrate for
our nanodisk direct numerical solution. 
The energy balance, in dimensional form, reads
\begin{equation}
 \frac{\partial}{\partial t}\left[E_k+E_s\right]+D=0,\label{eq:energy-sup}
\end{equation}
where we neglect the gravitational energy.  
Using the velocity field, Eqs.~(\ref{eq:vy}) and (\ref{eq:vx}) in Eqs.~(\ref{eq:kinetic}) and (\ref{eq:dissipation}),
the volume conservation $R(t)^2h(t)=R_0^2h_0$, and taking the time derivative in Eq.~(\ref{eq:energy-sup}), we
arrive at the following nonlinear second-order
variable-coefficient ordinary differential equation
\begin{equation}
 A(R)\ddot R + B(R)\dot R^2 + C(R)\dot R + D(R) = 0\, ,
 \label{eq:ODE2}
\end{equation}
where
$$
A(R)=\rho\left(\frac{4}{3}\frac{R_0^4h_0^2}{R^6}+\frac{1}{2}\right),\quad
B(R)=-\rho\left(\frac{4R_0^4h_0^2}{R^7}\right),\quad
$$
$$
C(R)=\frac{12\eta}{R^2},\quad
D(R)=\sigma\left(\frac{2 R}{R_0^2h_0}(1-\cos\theta_{\text{eq}}) - \frac{2}{R^2}\right),
$$
with the initial conditions $R(t=0)=R_0$ and $\dot R(t=0)=0$.
We compute $R(t)$ and $\dot R(t)$ by solving the initial value problem (\ref{eq:ODE2})
using the forth-order Runge-Kutta method. One would naturally consider detachment
when $R = 0$. Eq.~(\ref{eq:ODE2}) however becomes singular when $R\rightarrow 0$. 
We propose the following procedure to alleviate this numerical difficulty.
When detachment occurs, as $t\rightarrow\infty$, the total surface energy
will be equal to the surface energy of a spherical droplet
of radius $R_s=(\frac{3}{4}R_0^2h_0)^{1/3}$ (see Fig. \ref{fig:1}), thus
\begin{eqnarray}
 \sigma\left[\pi R(t\rightarrow\infty)^2(1-\cos\theta_{\text{eq}}) + 2\pi \frac{R_0^2h_0}{R(t\rightarrow\infty)}\right]  \nonumber &=& \\ 
 4\pi\sigma \left(\frac{3}{4}R_0^2h_0\right)^{2/3},
\label{eq:r-theta}
\end{eqnarray}
where $R(t\rightarrow\infty)$ is the radius of the wetted base as $t\rightarrow\infty$. 
We use Eq.~(\ref{eq:r-theta}) to calculate $R(t\rightarrow\infty)$ as a function of $\theta_{\text{eq}}$. 
Let us denote this detachment radius $R_d$.
When numerically computed $R(t)$ equals $R_d$, we consider the drop detached
and calculate its ejection velocity.  If $R(t)$ never reaches $R_d$, 
we then consider the disk only collapsing into a sessile drop on the substrate.
We note that Eq.~(\ref{eq:r-theta}) leads to a real positive solution for $R_d$ only for 
$\theta_{\text{eq}} \lesssim 107^{\circ}$,  where $R \lesssim 0.44$.   We  use this value 
of $R$ as the one at which the drop detaches from the substrate when $\theta_{\text{eq}} \ge 107^{\circ}$.

\begin{table}[t]
\begin{center}
\begin{tabular}{lllll}
\\[3pt]
\hline 
&Cu &Au\\
\hline 
$R_0$ (Cu), $L_0$(Au)		    &150 \AA           		&405 nm\\
$L_x, L_y, L_z$			    &375 \AA                   	&494 nm\\
$\Delta$	          	    &2.93 \AA                  	&1.5 nm\\
$\rho_{\text{air}}$(kg~m$^{-3}$)   	   &1.225             	&1.225\\
$\eta_{\text{air}}$(kg~m$^{-1}$s$^{-1}$)   &0.00002           	&0.00002\\
$\rho_{\text{drop}}$(kg~m$^{-3}$)          &7900              	&17310\\
$\eta_{\text{drop}}$(kg~m$^{-1}$s$^{-1}$)  &0.004288 ($@$ 1500 K)  &0.00425 ($@$ 1500 K)\\
$\sigma$(kg~s$^{-2}$)               &1.305                     	&1.15\\
$Oh=\eta/\sqrt{\rho\sigma L}$       &0.35                   	&0.047
\\
\hline 
\end{tabular}
\caption{Overview of the sets of parameters used in the numerical simulations \cite{Fuentes11,habenicht_05-sup}.
Initial conditions are specified in terms of disk radius, $R_0$ for Cu and a typical length of a 
triangle side, $L_0$ for Au.    In computing Ohnesorge number, $Oh$, for the typical length scale, 
$L$, we use $R_0$ for Cu and $L_0$ for Au.}
\label{table1}
\end{center}
\end{table}   

\section {Computational parameters}
Table~\ref{table1} provides an overview of the parameter sets used.
The computational domain is $L_x \times L_y \times L_z$ and
the maximum grid resolution of the adaptive mesh is represented by $\Delta$. 
For all the simulations, an open boundary condition (pressure and velocity gradient equal zero)
is imposed at the top and a symmetry boundary condition is imposed on the lateral sides.

\section{Retraction of a disk}
\begin{figure}[h]
\centering
\includegraphics[scale=1.5]{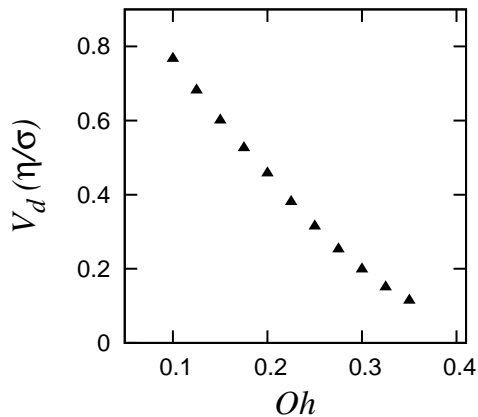}
\caption{Droplet detachment velocity $V_d$ nondimensionalized with the capillary characteristic 
velocity,  $\sigma/\eta$, as a function of the Ohnesorge number $Oh=\eta/\sqrt{\rho\sigma R_0}$ when
$\theta_\text{eq}=115^\circ$.}
\label{fig:model-sup}
\end{figure}

Figure~\ref{fig:model-sup} shows the results of additional investigation of the effective model (Eq.~(\ref{eq:ODE2})).   
Here, we plot the dimensionless ejection velocity, $V_d(\eta/\sigma)$, as
a function of $Oh$ when $\theta_\text{eq}=115^\circ$. The results show that the ejection velocity
increases approximately linearly as $Oh$ decreases. Increasing $Oh$ beyond a critical
value will result in no ejection. When $\theta_\text{eq}=80^\circ$ (figure not shown for brevity),
the trend of the results is consistent with the one shown in Fig.~\ref{fig:model-sup}, however, the range of $Oh$ for which a
detachment occurs is much smaller.

\section {Determination of the appropriate slip length}
\begin{figure}[h]
\centering
\includegraphics[scale=1.5]{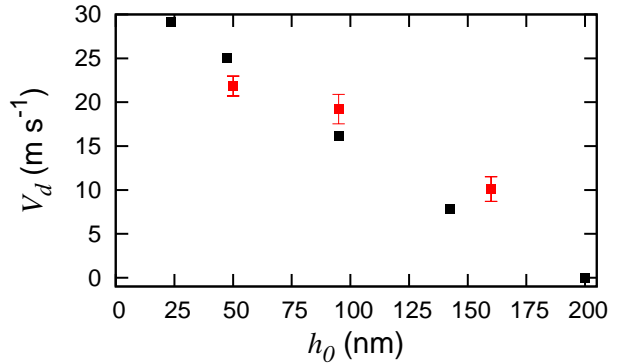}
\caption{Droplet detachment velocity $V_d$ for liquefied triangular Au nanostructures
with a side length of $405$ nm and various thicknesses, numerical results ({\small $\blacksquare$})
and experimental measurements ({\color[rgb]{1.00,0.00,0.00}{\small $\blacksquare$}}).
Numerical results are obtained using the slip length of $3$ nm.
}
\label{fig:Habenicht}
\end{figure}

We explored the influence of the slip length by carrying out
numerical experiments for Au structures.   Larger slip lengths
lead to faster collapse and large detachment velocity.   
In Fig.~\ref{fig:Habenicht}, we show that by choosing the slip length
of $3$ nm, the computed detachment velocity of nanostructures 
compares well with the experimental results from \cite{habenicht_05-sup}; this
value is therefore used in all presented Au simulations.

\end{document}